\documentclass[10pt,aps,prd,twocolumn,showpacs,superscriptaddress,longbibliography,nofootinbib]{revtex4-2}
\usepackage{mathrsfs,amsmath,amsthm,latexsym,amssymb,amsfonts,epsfig,cancel,enumerate,graphicx,txfonts,diagbox}

\usepackage{multirow}
\usepackage[T1]{fontenc}
\usepackage[utf8]{inputenc}
\usepackage[colorlinks=true,linkcolor=blue,citecolor=blue,urlcolor=blue]{hyperref}
\usepackage{orcidlink}

\usepackage{xcolor}
\usepackage{booktabs}
\usepackage{graphicx}
\definecolor{navy}{RGB}{0,0,150}
\usepackage{appendix}
\allowdisplaybreaks
\newcommand{\RGU}{Department of Physics, The Assam Royal Global University, Guwahati-781035, Assam, India}
\newcommand{\IMS}{The Institute of Mathematical Sciences, C. I. T. Campus, Taramani Chennai, 600113, India}
\newcommand{\UFM}{Programa de P\'os-Gradua\c c\~ao em F\'{\i}sica \& Coordena\c c\~ao do Curso de F\'{\i}sica -- Bacharelado, Universidade Federal do Maranh\~{a}o, 65085-580 S\~{a}o Lu\'{\i}s, Maranh\~{a}o, Brazil}

\begin{document}

\title{Charged Black Holes in Bumblebee gravity with Global Monopole: Thermodynamics and Shadow}

\author{Faizuddin Ahmed\orcidlink{0000-0003-2196-9622}}
\email{faizuddinahmed15@gmail.com}
\affiliation{\RGU}

\author{Shubham Kala\orcidlink{0000-0003-2379-0204}}
\email{shubhamkala871@gmail.com}
\affiliation{\IMS}

\author{Edilberto O. Silva\orcidlink{0000-0002-0297-5747}}
\email{edilberto.silva@ufma.br}
\affiliation{\UFM}

\date{\today}

\begin{abstract}
In this paper, we perform a detailed study of the thermodynamic properties of a charged black hole in bumblebee gravity in the presence of a global monopole. We also analyze the optical characteristics of this black hole solution, highlighting the influence of Lorentz symmetry violation and the global monopole on the black hole shadow. Furthermore, we examine the trajectories of both photons and test particles in this spacetime, showing how the geometric parameters alter their paths. Moreover, we study the dynamics of neutral test particles, with particular attention to the location of the innermost stable circular orbits (ISCOs). Finally, we investigate massless scalar perturbations and derive bounds on the greybody factors, illustrating how the black hole's geometric parameters affect field propagation, energy emission, and radiation sparsity in this background.
\end{abstract}

\maketitle

\section{Introduction}

The optical appearance of compact objects, shaped by strong gravitational lensing, photon capture, and the formation of bright rings around dark central regions, has become a central tool for probing gravity in the strong-field regime \cite{Synge1966,Luminet1979}. In particular, after the Event Horizon Telescope (EHT) observations of M87* \cite{EHTL1,EHTL4,EHTL6,EHTPRD2021}, and Sagittarius A* \cite{EHTL12,EHTL16,EHTL17}, black hole shadows have developed from a mainly theoretical subject into a mature observational framework. In recent years, this framework has been increasingly used not only to infer black hole parameters, but also to test the underlying gravitational theory itself \cite{Perlick2022,Wang2019,Vagnozzi2023}.

Shadow observables are especially valuable because they encode information about the photon sphere and, therefore, about the spacetime geometry near the compact object. This makes them useful for extracting intrinsic black hole parameters \cite{Hioki2009,Liu2025}, testing the no-hair paradigm \cite{Johannsen2010,Broderick2014}, and constraining departures from general relativity \cite{Vagnozzi2023}. Motivated by these developments, the literature has been extended far beyond standard Kerr or Reissner--Nordstr\"om black holes to include exotic compact objects and nonstandard geometries, such as naked singularities \cite{Shaikh2019}, gravastars \cite{Rosa2024}, and wormholes \cite{Cheng2026}. At the same time, recent studies have shown that modified gravity models may leave distinct signatures not only in shadow size, but also in lensing, orbital motion, and wave propagation.

Black hole physics is also deeply connected with thermodynamics. Beginning with Bekenstein's identification of black hole entropy with the horizon area \cite{Bekenstein1973} and Hawking's discovery that black holes radiate thermally \cite{Hawking1975}, black holes came to be understood as genuine thermodynamic systems. Their phase structure, local stability, and heat capacity provide important information about the interplay between geometry and quantum effects. In this context, the Hawking--Page transition and related thermal phenomena have remained fundamental in understanding how black holes exchange energy with their surroundings \cite{Page1983,Page2005}. For modified or topologically nontrivial spacetimes, thermodynamic observables often provide one of the clearest diagnostics of deviations from standard general relativity.

Another line of interest comes from topological defects that may have formed in the early universe. Among these, the global monopole is particularly relevant because it introduces a solid-angle deficit into spacetime, thereby changing the global geometry without necessarily altering the basic spherical symmetry of the solution \cite{Kibble1976,Vilenkin1985,Barriola1989}. Black holes endowed with global monopoles have therefore been widely studied as laboratories for investigating how nontrivial topology affects lensing, particle dynamics, thermodynamics, and other observable properties. The conical asymptotics generated by a global monopole also make these systems conceptually interesting, since they differ in an essential way from asymptotically flat black holes.

Lorentz symmetry violation has garnered significant attention in various effective and phenomenological approaches to gravity. A prominent framework in this context is \emph{Bumblebee gravity}, in which Lorentz symmetry is spontaneously broken by a vector field acquiring a nonzero vacuum expectation value. This mechanism alters the spacetime geometry and can give rise to black hole solutions with nontrivial physical and geometric features \cite{Casana2018,Li2026,Maluf2021}. Black hole solutions in Bumblebee gravity incorporating topological defects, such as global monopoles, have been studied in a few works \cite{Gullu2021,Gogoi2022}. More recently, non-rotating and rotating Bumblebee black holes have been explored, along with their properties including thermodynamics, greybody factors, accretion processes, and quasinormal modes \cite{Gao2024,Ding2025,Islam2024JCAP, AzregAinou2025,Lambiase2023,Liu2025arxiv,Capozziello2023,Ahmed2026arxiv}. The combination of Lorentz-violating effec, global monopoles, and electric charge provides a rich framework for investigatingte how both local and global geometric deformations influence observable black hole phenomena.

Although astrophysical black holes are usually expected to be nearly neutral, charged black holes continue to play a significant theoretical role. Electric charge provides an additional geometric scale and changes the structure of horizons, circular orbits, and wave propagation. In more speculative settings, such as primordial or microscopic black holes, a non-negligible charge may also be phenomenologically relevant. For these reasons, charged black hole solutions remain important testing grounds for the combined effects of modified gravity, topological defects, and electromagnetic structure.

Motivated by all these developments, in this paper, we investigate a charged black hole in bumblebee gravity in the presence of a global monopole. We analyze its thermodynamic properties, including the Hawking temperature, Gibbs free energy, and specific heat. We then examine the photon sphere and the corresponding shadow, followed by the weak-field deflection of light and the motion of massive test particles, including perihelion precession and circular-orbit dynamics. Finally, we study massless scalar perturbations, derive bounds on the greybody factors, and discuss the associated energy emission rate and the sparsity of Hawking radiation. Altogether, these observables provide a coherent picture of how Lorentz-violating effects, electric charge, and topological defects modify the physical behavior of black holes in the strong-gravity regime.

\section{Spacetime metric}\label{sec2}

A charged black hole solution in the bumblebee gravity frame can be expressed as \cite{Liu2025EPJC,Li2026}
\begin{equation}
\mathrm{d} s^2=-h(r)dt^{2}+\frac{1+\ell}{h(r)}dr^{2}+r^{2}(d\theta ^{2}+\sin^{2} \theta \,d\phi^2),\label{aa1}
\end{equation}
where
\begin{align}
h(r)=1-\frac{2m}{r}+\frac{2(1+\ell)\,q^{2}}{(2+\ell)\,r^{2}}, \label{aa2}
\end{align}
where $\ell=\xi b^2$ is the Lorentz-violating (LV) parameter and \(q\) is the charge parameter.

For the metric (\ref{aa1}), setting the Lorentz-violation (LV) parameter to zero, $\ell = 0$, the spacetime reduces to the Reissner--Nordstr\"om black hole solution. In the absence of electric charge, i.e., $q = 0$, the geometry describes a Schwarzschild-like black hole in bumblebee gravity~\cite{Casana2018}. Furthermore, when both parameters vanish, $\ell = 0$ and $q = 0$, the solution consistently recovers the standard Schwarzschild black hole.

Now, we want to include a global monopole into the black hole solution. A basic model to describe a global monopole involves a triplet of scalar fields with a global \(O(3)\) symmetry that is spontaneously broken into \(U(1)\) symmetry. It is worth noting that the gravitational field of a global monopole yields a spacetime with a solid-angle deficit. Thereby, the Lagrangian density that describes the global monopole reads
\begin{equation}
     {\mathcal {L}}_m=-\frac{1}{2}\partial _\mu \psi ^a\partial ^\mu \psi^{*\,a} - \frac{\chi }{4}\left(\psi ^a\psi^{*\,a}-\eta ^2_0\right)^2, \label{aa3}
\end{equation}
where $\psi^a$ is a multiplet of scalar fields, $\chi$ is a coupling constant and $\eta_0$ is the energy scale of the symmetry breaking. Above, the index \(a\) indicates the labels of the scalar fields and runs from 1 to 3. To proceed, we introduce the following ansatz for describing the monopole:
\begin{equation}
    \psi ^a=\eta_0 \frac{x^a}{r},\label{aa4}
\end{equation}
where $x^{a}\,x^{a}=r^2$. It is worth noting that this ansatz is approximated outside the GM core. The energy-momentum tensor components are given by \cite{Barriola1989}
\begin{equation}
    T^{t\,(\rm GM)}_{t}=\frac{\eta^2_0}{r^2}=T^{r\,(\rm GM)}_{r},\quad T^{\theta\,(\rm GM)}_{\theta}=T^{\phi\,(\rm GM)}_{\phi}= 0.\label{aa5}
\end{equation}

The presence of a global monopole modifies the black hole spacetime by increasing the radius of the event horizon and introducing a conical deficit \cite{Barriola1989}. These modifications can have significant astrophysical consequences, influencing particle dynamics, accretion processes, and observational signatures, thereby motivating further investigation into the role and impact of a global monopole in strong gravitational fields.

Therefore, a static spherically symmetric metric of a charged black hole with a global monopole in bumblebee gravity is described by
\begin{equation}
\tilde{\mathrm{d} s}^2=-f(\tilde{r})d\tilde{t}^{2}+\frac{1+\ell}{f(\tilde{r})}d\tilde{r}^{2}+\tilde{r}^{2}(d\theta ^{2}+\sin^{2} \theta \,d\phi^2),\label{metric0}
\end{equation}
where
\begin{align}
f(\tilde{r})=1-8\pi \eta^2_0-\frac{2\tilde{m}}{\tilde{r}}+\lambda\,\frac{\tilde{q}^{2}}{\tilde{r}^{2}},\quad \lambda=\frac{1+\ell}{1+\ell/2}. \label{function0}
\end{align}
Here $\tilde{m}$ and $\tilde{q}$ are the mass and electric charge parameters, respectively.

We make the following coordinate transformations:
\begin{align}
\tilde{t}=t/\beta,\qquad \tilde{r}=r \beta,\qquad \beta^2=1-8\pi \eta^2_0\label{trans-1}
\end{align}
and introduce the new parameters
\begin{align}
    m=\tilde{m}/\beta^3, \qquad q=\tilde{q}/\beta^2 ,\qquad \eta^2=8\pi \eta^2_0.\label{tarns-2}
\end{align}

Finally, we arrive at the following line element:
\begin{equation}
\mathrm{d} s^2=-f(r)dt^{2}+\frac{1+\ell}{f(r)}dr^{2}+\beta^2\,r^{2}(d\theta ^{2}+\sin^{2} \theta\, d\phi^2),\label{metric}
\end{equation}
where $\beta^2=1-\eta^2$ and
\begin{align}
f(r)=1-\frac{2m}{r}+\lambda\,\frac{q^{2}}{r^{2}}. \label{function}
\end{align}
Hence, the spacetime described by this metric (\ref{metric}) exhibits a solid angle deficit. Also, the physical parameters, the black hole charge $Q$, and the ADM mass $M$ are given by \cite{Liu2025EPJC,Deng2018,Rizwan2018}
\begin{align}
    Q&=-\frac{1}{4\pi }\int _{\partial \Sigma }d\theta d\phi \sqrt{\gamma ^{(2)}}n_{\mu }\sigma _{\nu } \left( F^{\mu \nu }+\frac{\xi }{l +2} B^{\mu \nu }B^{\alpha \beta }F_{\alpha \beta }\right) \nonumber \\
    &=\beta^2\,\left( 1+b^{2}\frac{\xi }{(l +2)}\right) q=\beta^2 \frac{2(1+\ell )}{2+\ell }=\lambda\,\beta^2\,q,\nonumber\\ 
   M&=\frac{1}{8 \pi} \oint \xi^{\mu;\nu}_{(t)} d^2\Sigma_{\mu\nu}=\frac{\beta^2 m}{\sqrt{1+\ell}}.\label{parameters}
\end{align}

Here, $\Sigma$ denotes a three-dimensional spacelike hypersurface endowed with the induced metric $\gamma_{ij}^{(3)}$. Its boundary, $\partial \Sigma$, is a two-sphere located at spatial infinity, with the induced metric $\gamma_{ij}^{(2)} = \beta^2\, r^{2} \, d\Omega^{2}$. The vectors $n_\mu = (1,0,0,0)$ and $\sigma_\mu = (0,1,0,0)$ represent the unit normal vectors associated with $\partial \Sigma$ and $\Sigma$, respectively.

\begin{figure*}[tbhp]
\centerline{
\includegraphics[width=80mm,height=70mm]{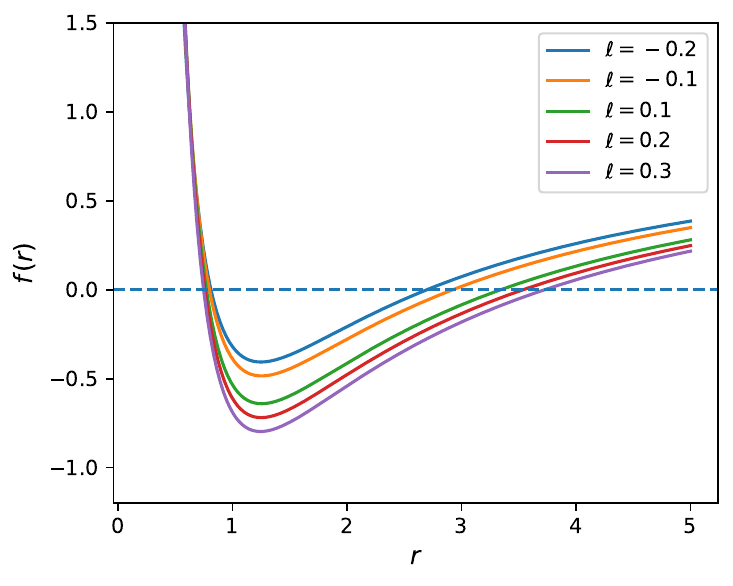}\qquad
\includegraphics[width=80mm,height=70mm]{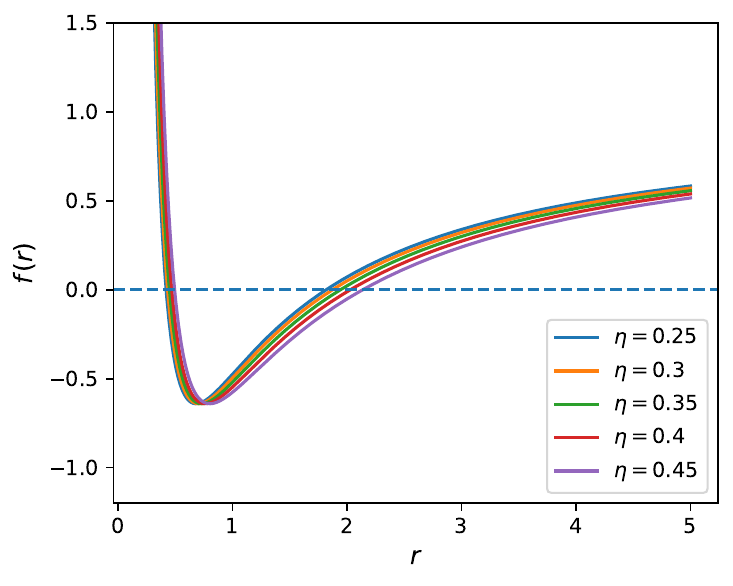}}(a) $\eta=0.3$ \hspace{5cm} (b) $\ell=0.2$
\caption{Metric function $f(r)$ versus the radial coordinate $r$ for fixed charge $Q=0.8$. Panel (a) shows the effect of varying the Lorentz-violating parameter $\ell$ at fixed monopole parameter $\eta=0.3$, while panel (b) shows the effect of varying $\eta$ at fixed $\ell=0.2$. The zeros of each curve correspond to the inner and outer horizons.}
\label{fig:metric-function}
\end{figure*}

The metric function $f(r)$ in terms of $Q$ can be re-written as
\begin{align}
f(r)=1-\frac{\sqrt{1+\ell}}{\beta^2}\,\frac{2M}{r}+\frac{1}{\lambda \beta^4}\frac{Q^{2}}{r^{2}}=1-\frac{2 M \zeta}{r}+\frac{\varsigma Q^2}{r^2},\label{function-2}
\end{align}
where
\begin{equation}
\zeta=\frac{\sqrt{1+\ell}}{\beta^2},\quad 
\varsigma=\frac{1}{\lambda \beta^4}.\label{fucntion-3}
\end{equation}

In the limit $\ell=0$, the considered spacetime (\ref{metric}) reduces to the RN black hole with a global monopole in Einstein-Maxwell gravity, and is given by ($d\Omega^2=d\theta ^{2}+\sin^{2} \theta\,d\phi^2$) \cite{Deng2018,Rizwan2018}
\begin{align}
\mathrm{d} s^2&=-F(r)dt^{2}+\frac{1}{F(r)}dr^{2}+(1-\eta^2)\,r^{2} d\Omega^2,\nonumber\\
F(r)&=1-\frac{2M}{(1-\eta^2) r}+\frac{Q^{2}}{\left(1-\eta^2\right)^2 r^{2}}.
\end{align}

Fig.~\ref{fig:metric-function} shows the variation of the metric function $f(r)$ with radial distance $r$ for different values of the LV parameter $\ell$ and GM parameter $\eta$, with $Q=0.8$. It is observed that, for the chosen values of both parameters, the black hole possesses two horizons. The global monopole field effectively introduces a deficit solid angle, thereby modifying the asymptotic structure of spacetime and influencing the radial profile of the metric function.

\section{Thermodynamics}

In this section, we investigate the thermodynamic properties of the system and analyze how Lorentz violation and the global monopole modify these properties. Several works have explored the thermodynamic properties of black holes in various configurations (see \cite{FA1,FA2,FA3} and references therein).

The largest root of $f(r_h)=0$ gives the black hole event horizon, where $r_h$ is the location of the horizon. One can use this to express the ADM mass as follows:
\begin{equation}
M=\frac{r_h}{2\zeta}+\frac{\varsigma Q^2}{2 r_h \zeta}.\label{cc1}
\end{equation}
Moreover, the horizon area is obtained as
\begin{equation}
\mathcal{A}=\lim_{r \to r_h} \int\sqrt{g_{\theta\theta}\,g_{\phi\phi}}\, d\theta\, d\phi.\label{cc2}
\end{equation}
In our case, we find
\begin{equation}
\mathcal{A}=4\pi \beta^2\,r_h^2.\label{cc3}
\end{equation}
The entropy of the black hole from the horizon area is
\begin{equation}
S=\mathcal{A}/4=\pi \beta^2\,r_h^2.\label{cc4}
\end{equation}

The surface gravity can be determined from the relation $\kappa^2=-\frac{1}{2} \xi_{\mu;\nu}\,\xi^{\mu;\nu}|_{r=r_h}$ \cite{Hawking1975,Bekenstein1973}, where $\xi^{\mu}$ is the timelike Killing vector $(\partial_t)^{\mu}$. This is given by
\begin{equation}
    \kappa=\frac{f'(r_h)}{2 \sqrt{1+\ell}}=\frac{1}{2 r_h \sqrt{1+\ell}}\left(1-\varsigma\,\frac{Q^2}{r_h^2}\right).\label{cc5}
\end{equation}

Therefore, the Hawking temperature is given by
\begin{align}
    T_H=\frac{\kappa}{2\pi}&=\frac{1}{4 \pi r_h \sqrt{1+\ell}}\left(1-\varsigma\,\frac{Q^2}{r_h^2}\right)\nonumber\\
    &=\frac{1}{4 \pi r_h \sqrt{1+\ell}}\left(1-\frac{2+\ell}{2(1+\ell)(1-\eta^2)^2}\,\frac{Q^2}{r_h^2}\right).\label{cc6}
\end{align}
In the limit $\ell=0$, the Hawking temperature simplifies to those results obtained in \cite{Deng2018,Rizwan2018}.

\begin{figure*}[tbhp]
\centerline{
\includegraphics[width=80mm,height=70mm]{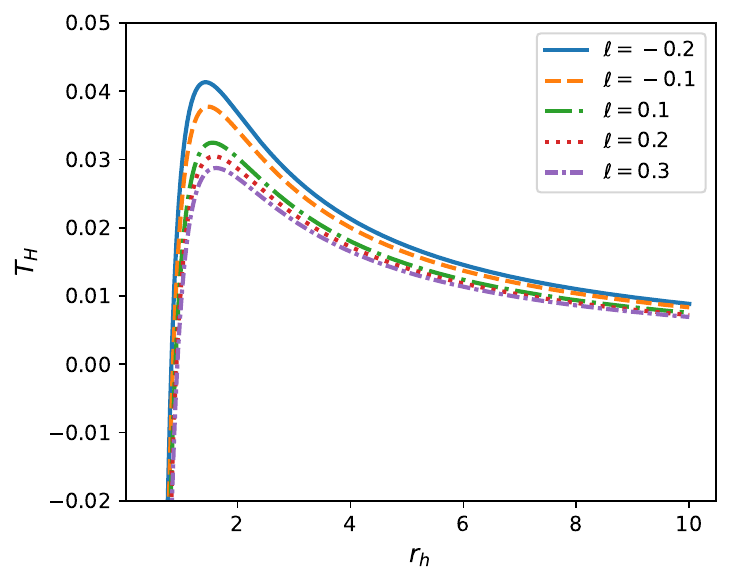}\qquad
\includegraphics[width=80mm,height=70mm]{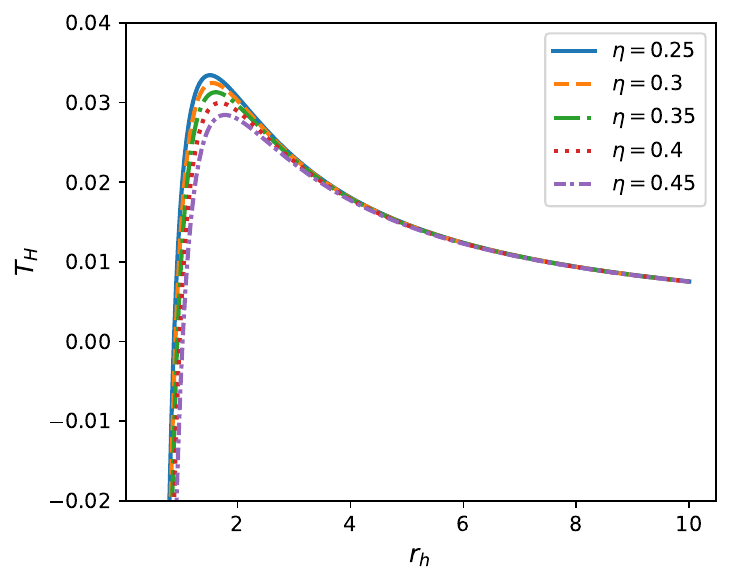}}(a) $\eta=0.3$ \hspace{5cm} (b) $\ell=0.2$
\caption{Hawking temperature $T_H$ as a function of the horizon radius $r_h$ for fixed $Q=0.8$. Panel (a) shows the dependence on $\ell$ at fixed $\eta=0.3$, while panel (b) shows the dependence on $\eta$ at fixed $\ell=0.2$. The curves display the typical non-monotonic behavior of charged black holes, with a maximum separating different thermal regimes.}
\label{fig:Hawking}
\end{figure*}

In Fig.~\ref{fig:Hawking}, we plot the variation of the Hawking temperature $T_H$ as a function of the horizon radius $r_h$ for different values of the LV parameter $\ell$ and the GM parameter $\eta$, keeping the charge fixed at $Q=0.8$. In the left panel, we observe that an increase in the LV parameter reduces the peak Hawking temperature and slightly shifts its position. This suggests that the effect of the LV parameter on the Hawking temperature is to diminish the black hole's ability to emit radiation. In the right panel, we find that the effect of the GM parameter on the Hawking temperature is relatively weak, with the peak value decreasing slightly. In both cases, the temperature exhibits a non-monotonic behavior, increasing rapidly for small $r_h$, reaching a maximum, and then gradually decreasing for larger horizon radii. This behavior reflects the interplay among gravitational, electromagnetic, and topological effects in determining the black hole's thermodynamic properties.

\begin{figure*}[tbhp]
	\centerline{
		\includegraphics[width=80mm,height=70mm]{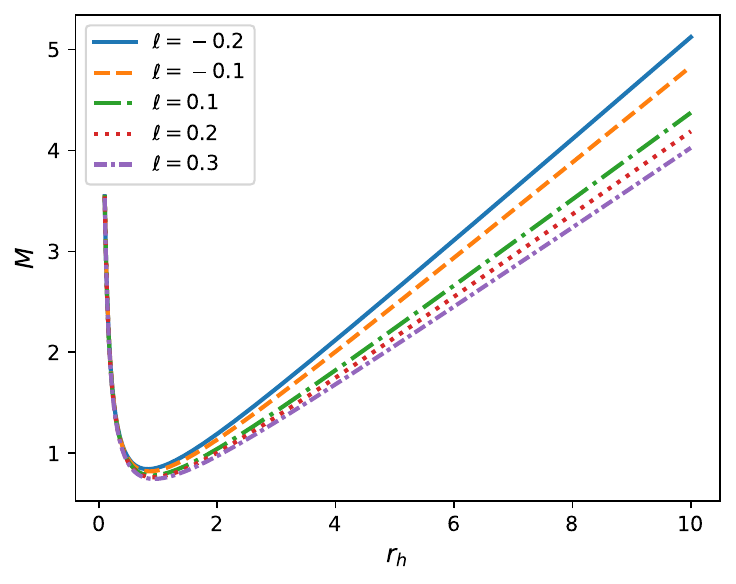}\qquad
        \includegraphics[width=80mm,height=70mm]{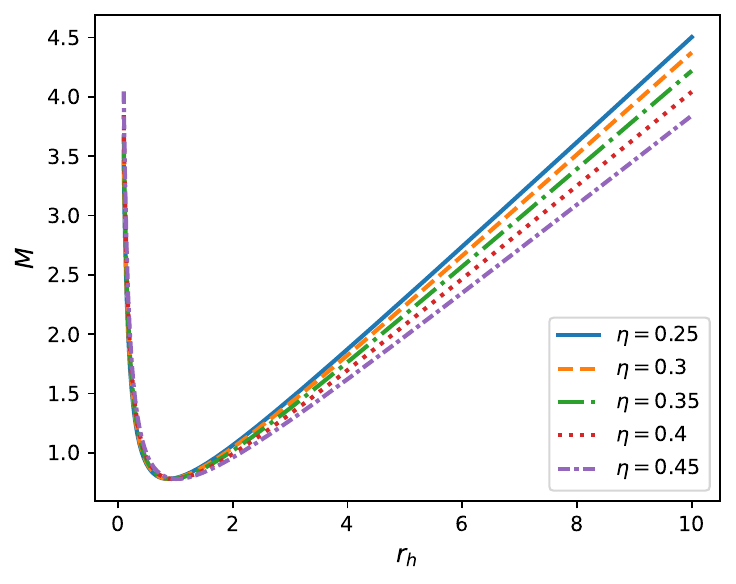}}(a) $\eta=0.3$ \hspace{5cm} (b) $\ell=0.2$
	\caption{ADM mass $M$ as a function of the horizon radius $r_h$ for fixed $Q=0.8$. Panel (a) corresponds to $\eta=0.3$ with varying $\ell$, and panel (b) corresponds to $\ell=0.2$ with varying $\eta$. The curves show how the mass--radius relation is deformed by Lorentz violation and the monopole background.}
\label{fig:ADMmass}
\end{figure*}

\begin{figure*}[tbhp]
	\centerline{
		\includegraphics[width=80mm,height=70mm]{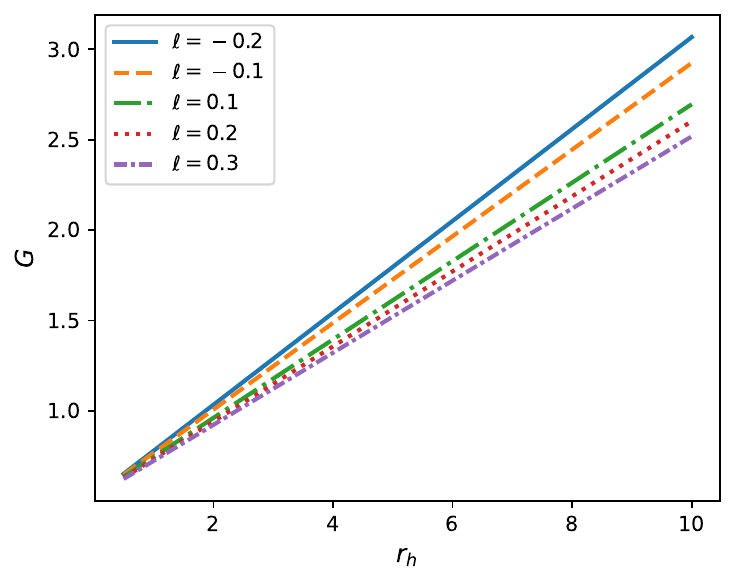}\qquad
        \includegraphics[width=80mm,height=70mm]{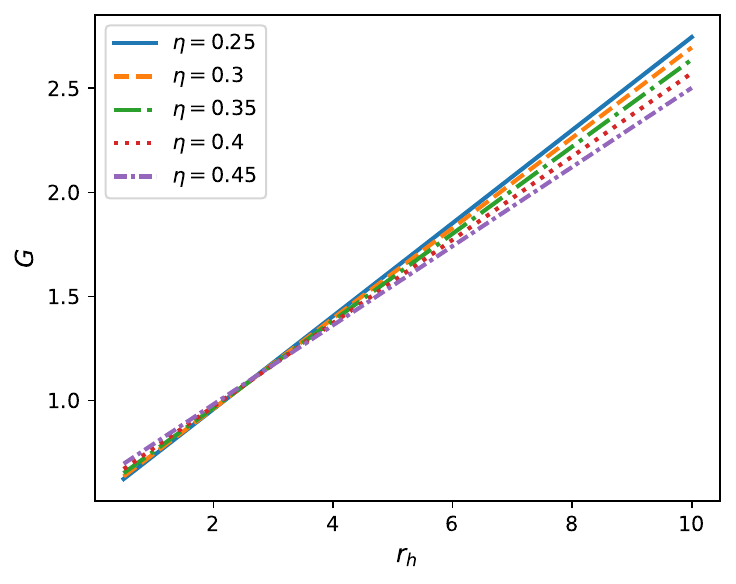}}(a) $\eta=0.3$ \hspace{5cm} (b) $\ell=0.2$
	\caption{Gibbs free energy $G$ as a function of the horizon radius $r_h$ for fixed $Q=0.8$. Panel (a) shows varying $\ell$ at fixed $\eta=0.3$, while panel (b) shows varying $\eta$ at fixed $\ell=0.2$. The monotonic behavior indicates how the thermodynamic potential shifts with the geometric parameters.}
\label{fig:Gibbs}
\end{figure*}

\begin{figure*}[tbhp]
	\centerline{
		\includegraphics[width=80mm,height=70mm]{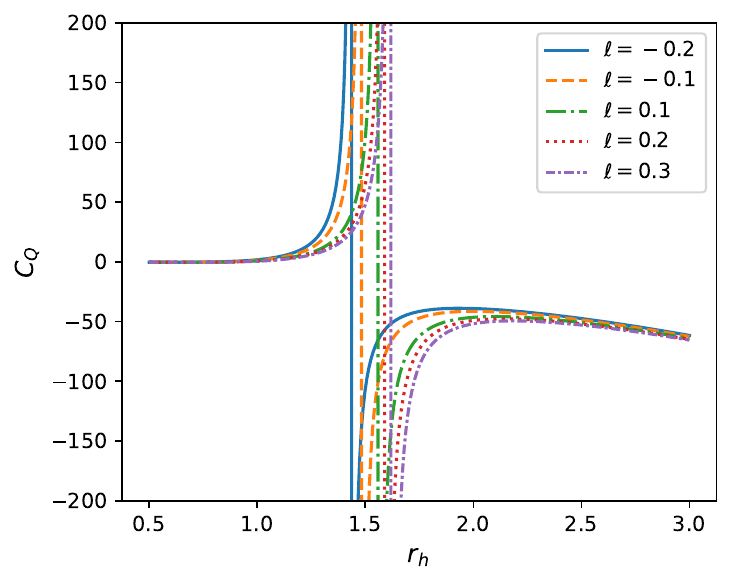}\qquad
        \includegraphics[width=80mm,height=70mm]{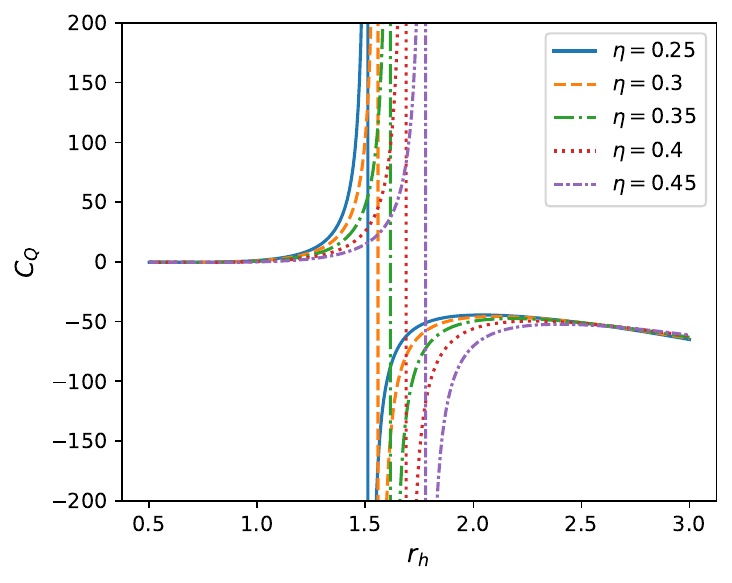}}(a) $\eta=0.3$ \hspace{5cm} (b) $\ell=0.2$
	\caption{Specific heat at constant charge, $C_Q$, as a function of the horizon radius $r_h$ for fixed $Q=0.8$. Panel (a) corresponds to $\eta=0.3$ with varying $\ell$, and panel (b) corresponds to $\ell=0.2$ with varying $\eta$. The divergence marks the transition between locally stable ($C_Q>0$) and unstable ($C_Q<0$) branches.}
\label{fig:SpecificHeat}
\end{figure*}

Expressing the ADM mass \(M\) in terms of the entropy \(S\), we obtain
\begin{equation}
    M=\frac{1}{2 \zeta}\sqrt{\frac{S}{\pi}}+\frac{\varsigma\,Q^2}{2 \zeta}\sqrt{\frac{\pi}{S}}.\label{cc7}
\end{equation}
From the above, we see that the ADM mass $M=M(S, Q)$ and hence, the differential mass can be expressed as
\begin{equation}
    dM=\left(\frac{\partial M}{\partial S}\right)_{Q}\,dS+\left(\frac{\partial M}{\partial Q}\right)_{S}\,dQ=T\,dS+\Phi\,dQ,\label{cc8}
\end{equation}
where
\begin{align}
    T&=\left(\frac{\partial M}{\partial S}\right)_Q=\frac{1}{4 \zeta \sqrt{\pi S}}-\frac{\varsigma\,Q^2}{4 \zeta S}\sqrt{\frac{\pi}{S}}=T_H,\label{cc9}\\
    \Phi&=\left(\frac{\partial M}{\partial Q}\right)_{S}=\frac{\varsigma\,Q}{ \zeta}\sqrt{\frac{\pi}{S}}=\frac{2+\ell}{2(1+\ell)^{3/2}}\,\frac{1}{(1-\eta^2)}\,\frac{Q}{r_h}.\label{cc10}
\end{align}
Moreover, the thermodynamic system obeys the Smarr relation for this black hole, given by \cite{Smarr1973a,Smarr1973b}
\begin{equation}
    M=2 T S+Q\,\Phi.\label{cc11}
\end{equation}
Fig.~\ref{fig:ADMmass} depicts the dependence of the ADM mass of the black hole on the horizon radius while maintaining a fixed charge of $Q=0.8$. In both scenarios, the ADM mass first reaches a peak, then declines to a minimum near the horizon. After this minimum, the mass increases roughly in a linear manner as the horizon radius grows. Additionally, it is observed that larger values of $\ell$ and $\eta$ correspond to a reduction in the ADM mass. This trend suggests that these parameters contribute to a decrease in the system's gravitational energy, thereby altering the relationship between mass and radius and affecting the overall structure of spacetime. Furthermore, the Gibbs free energy can be obtained as \cite{Page1983,Page2005}
\begin{equation}
    G=M-TS=\frac{r_h (1-\eta^2)}{4\sqrt{1+\ell}}\left[1+\frac{3Q^2}{\lambda (1-\eta^2)^{3}\,r_h^2}\right].\label{cc12}
\end{equation}
Fig.~\ref{fig:Gibbs} illustrates how the Gibbs free energy varies with the horizon radius for different values of $\ell$ and $\eta$, with the charge fixed at $Q = 0.8$. In the left panel, $G$ increases monotonically as $r_h$ grows, and an increase in $\ell$ corresponds to a consistent reduction in the Gibbs free energy, suggesting that the Lorentz-violating effects tend to decrease the thermodynamic potential. The right panel shows a similar pattern with respect to $\eta$, though its influence appears less pronounced. The near-linear dependence of $G$ on $r_h$ implies a gradual thermodynamic change, with no clear evidence of a phase transition within the explored parameter range.

Finally, the specific heat capacity at constant charge is given by \cite{Page1983,Page2005,Bardeen1973}
\begin{equation}
C_Q=\left(\frac{\partial M}{\partial T}\right)_Q=T\,\left(\frac{\partial S}{\partial T}\right)_Q
=2\pi\beta^2\,
\frac{r_h^2\left(r_h^2-\frac{Q^2}{\lambda \beta^4}\right)}
{\frac{3 Q^2}{\lambda \beta^4}-r_h^2}.\label{cc13}
\end{equation}
Fig.~\ref{fig:SpecificHeat} presents the behavior of the specific heat at constant charge, $C_Q$, as a function of the horizon radius $r_h$ for various values of $\ell$ and $\eta$, with the charge fixed at $Q=0.8$. In both panels, $C_Q$ diverges at a certain critical horizon radius, which suggests the occurrence of a second-order phase transition. When $r_h$ is small, the specific heat remains positive, indicating thermodynamically stable states. As $r_h$ increases beyond the critical point, $C_Q$ becomes negative, reflecting an unstable regime. The location of this divergence varies depending on the chosen values of $\ell$ and $\eta$, implying that these parameters play a role in determining the thermal stability and phase characteristics of the black hole. Our analysis of the thermodynamic properties indicates that the Lorentz-violating parameter $\ell$ and the global monopole parameter $\eta$ jointly affect the thermodynamic quantities, resulting in deviations from those of a standard charged black hole.

\section{Photon sphere and Shadow}

We analyze the null geodesic motion in the equatorial plane, defined by $\theta = \pi/2$ and $\dot{\theta} = 0$. Imposing the null condition $ds^2 = 0$ for the metric (\ref{metric}), we obtain
\begin{equation}
    -f(r)\,\dot{t}^2 + \frac{1+\ell}{f(r)}\,\dot{r}^2 + \beta^2\,r^2\,\dot{\phi}^2 = 0.
    \label{dd1}
\end{equation}

Since the spacetime metric is independent of $t$ and $\phi$, there exist two conserved quantities associated with these symmetries: the conserved energy $\mathrm{E}$ and the conserved angular momentum $\mathrm{L}$. These are given by \cite{Galtsov1989,PLB2025}
\begin{equation}
    \dot{t} = \frac{\mathrm{E}}{f(r)}, 
    \qquad 
    \dot{\phi} = \frac{\mathrm{L}}{\beta r^2}.
    \label{dd2}
\end{equation}

Substituting Eq.~(\ref{dd2}) into Eq.~(\ref{dd1}), the radial equation of motion becomes
\begin{equation}
    \dot{r}^2 = V_{\rm eff},
    \label{dd3}
\end{equation}
where the effective potential is defined as
\begin{equation}
    V_{\rm eff}(r) = \frac{1}{1+\ell}\left(\mathrm{E}^2 - \frac{\mathrm{L}^2}{r^2}\,f(r)\right).
    \label{dd4}
\end{equation}

\begin{figure*}[tbhp]
	\centerline{
		\includegraphics[width=80mm,height=70mm]{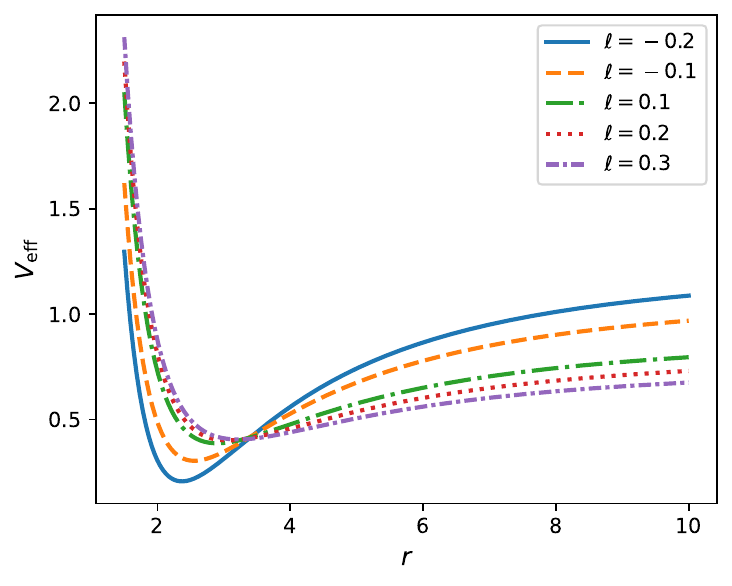}\qquad
        \includegraphics[width=80mm,height=70mm]{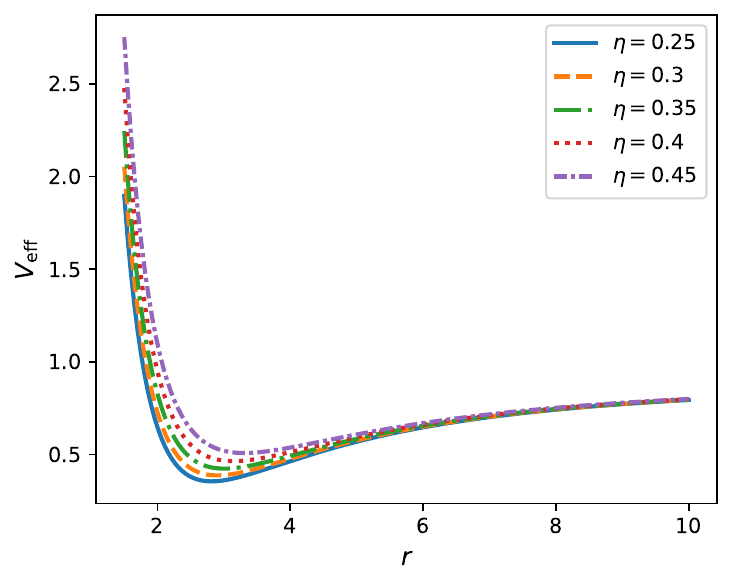}}(a) $\eta=0.3$ \hspace{5cm} (b) $\ell=0.2$
	\caption{Null effective potential $V_{\rm eff}(r)$ for fixed $L=4$ and $Q=0.8$. Panel (a) shows varying $\ell$ at fixed $\eta=0.3$, while panel (b) shows varying $\eta$ at fixed $\ell=0.2$. The extremum of the potential is associated with the unstable photon sphere.}
\label{fig:VEffective}
\end{figure*}

Fig.~\ref{fig:VEffective} presents the dependence of the null effective potential on the radial distance for various values of the parameters $\ell$ and $\eta$. The relevant extremum of this potential determines the unstable circular null orbit associated with the photon sphere. The profiles also show how the LV and GM parameters shift the potential structure, thereby modifying photon trajectories.

We now investigate circular photon orbits and the associated physical quantities. For circular orbits at $r = r_c$, the conditions $\dot{r} = 0$ and $\ddot{r} = 0$ must be satisfied. Using Eq.~(\ref{dd3}), these conditions lead to
\begin{equation}
    \mathrm{E}^2 = \frac{\mathrm{L}^2}{r^2}\,f(r),
    \qquad
    V'_{\rm eff}(r) = 0.
    \label{dd5}
\end{equation}

The first condition determines the critical impact parameter for photons, given by
\begin{equation}
    \frac{1}{b_c} = \frac{\mathrm{E}}{\mathrm{L}}
    =\frac{\sqrt{f(r_s)}}{r_s},
    \qquad\Longleftrightarrow\qquad
    b_c=\frac{r_s}{\sqrt{f(r_s)}}.
    \label{dd6}
\end{equation}

\begin{figure*}[tbhp]
	\centerline{
		\includegraphics[width=110mm,height=90mm]{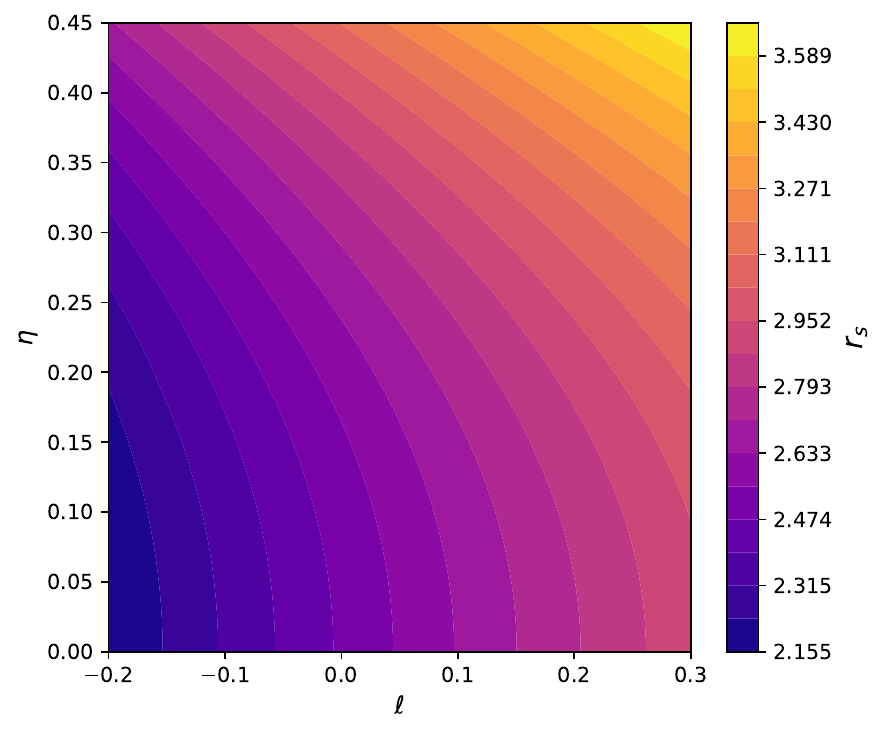}}    
	\caption{Two-dimensional map of the photon-sphere radius $r_s$ in the $(\ell,\eta)$ parameter plane for fixed $Q=0.8$. The color scale (or contour level) indicates how the size of the photon sphere changes jointly with the Lorentz-violating parameter and the monopole parameter.}
\label{fig:2Dphotonsphere}
\end{figure*}

The second condition, $V'_{\rm eff}(r) = 0$, determines the photon sphere radius $r = r_s$, which satisfies
\begin{equation}
    r_s^2 - 3M \zeta r_s + 2 \varsigma Q^2 = 0.\label{dd7}
\end{equation}
Solving this quadratic equation yields
\begin{equation}
   r_s = \frac{3M \zeta+ \sqrt{9M^2 \zeta^2- 8\varsigma Q^2}}{2}.
   \label{dd8}
\end{equation}
Fig.~\ref{fig:2Dphotonsphere} displays a two-dimensional map of the photon-sphere radius in the $(\ell,\eta)$ plane. The radius increases as either parameter grows, indicating that Lorentz violation and the monopole-induced conical deficit tend to move the unstable photon orbit outward.

The photon sphere exists provided that
\begin{equation}
  Q^2 \leq  \frac{9 M^2 \lambda (1+\ell)}{8}.
  \label{constraint}
\end{equation}

Next, we determine the shadow radius $R_{\rm sh}$ of the black hole. Since the considered spacetime is asymptotically conical at spatial infinity, the shadow radius for a static distant observer is given by
\begin{equation}
    R_{\rm sh} =\frac{r_s}{\sqrt{f(r_s)}}=\frac{r_s^2}{\sqrt{r_s^2-2 M \zeta r_s+\varsigma Q^2}}.
    \label{dd12}
\end{equation}

\begin{figure*}[tbhp]
	\centerline{
		\includegraphics[width=160mm,height=80mm]{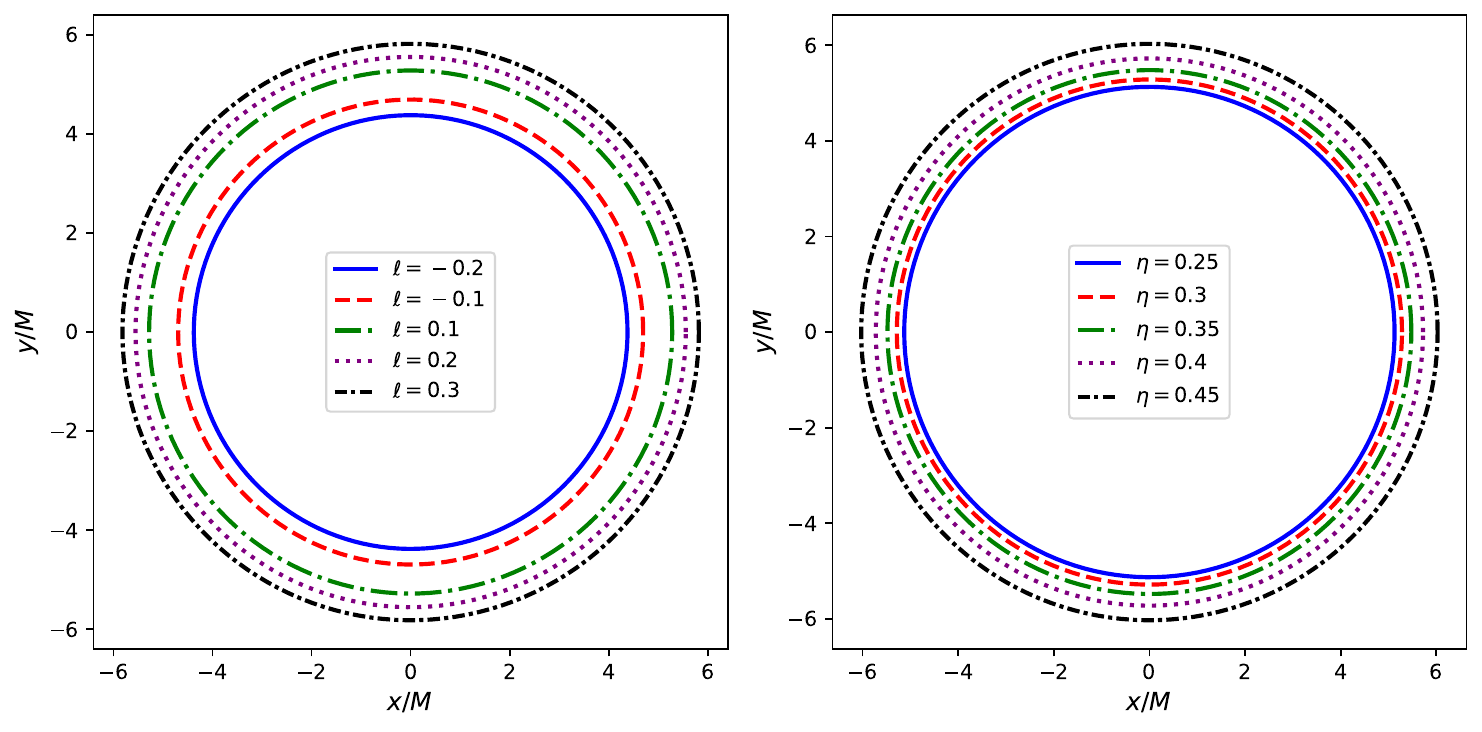}} (a) $\eta=0.3$ \hspace{5cm} (b) $\ell=0.2$   
	\caption{Parametric representation of the black hole shadow for different values of $\ell$ and $\eta$ at fixed $Q=0.8$. Panel (a) shows shadows for varying $\ell$ at fixed $\eta=0.3$, while panel (b) shows shadows for varying $\eta$ at fixed $\ell=0.2$. In both cases, the shadow remains circular, but its radius changes with the geometric parameters.}
\label{fig:ShadowPara}
\end{figure*}

\begin{table*}[tbhp]
\centering
\renewcommand{\arraystretch}{1.4}
\begin{tabular}{|c| c| c| c|}
\hline
Case & Conditions & $r_s$ & $R_{\rm sh}$ \\
\hline

(i) & $\ell=0,\ \eta\neq0$ 
& $\displaystyle \frac{3M+\sqrt{9M^2-8Q^2}}{2(1-\eta^2)}$
& $\displaystyle 
\frac{\sqrt{1-\eta^2}\,r_s^2}{\sqrt{r_s^2-\frac{2M}{1-\eta^2}r_s+\frac{Q^2}{(1-\eta^2)^2}}}$ \\
\hline

(ii) & $\eta=0,\ \ell\neq 0$ \cite{Li2026}
& $\displaystyle 
\frac{3M\sqrt{1+\ell}+\sqrt{9M^2(1+\ell)-8Q^2\frac{1+\ell}{1+\ell/2}}}{2}$
& $\displaystyle 
\frac{r_s^2}{\sqrt{r_s^2-2M\sqrt{1+\ell}\,r_s+\frac{1+\ell}{1+\ell/2}Q^2}}$ \\
\hline

(iii) & $\ell=0=\eta$ \cite{Eiroa2002}
& $\displaystyle \frac{3M+\sqrt{9M^2-8Q^2}}{2}$
& $\displaystyle 
\frac{r_s^2}{\sqrt{r_s^2-2Mr_s+Q^2}}$ \\
\hline

(iv) & $Q=0$ 
& $\displaystyle \frac{3M\sqrt{1+\ell}}{1-\eta^2}$
& $\displaystyle 
3\sqrt{3}M\,\frac{\sqrt{1+\ell}}{\sqrt{1-\eta^2}}$ \\
\hline

(v) & $Q=0=\ell$ \cite{Barriola1989}
& $\displaystyle \frac{3M}{1-\eta^2}$
& $\displaystyle 
\frac{3\sqrt{3}M}{\sqrt{1-\eta^2}}$ \\
\hline

(vi) & $Q=0=\eta$ \cite{Casana2018}
& $\displaystyle 3M\sqrt{1+\ell}$
& $\displaystyle 
3\sqrt{3}M\,\sqrt{1+\ell}$ \\
\hline

\end{tabular}
\caption{Photon-sphere radius $r_s$ and shadow radius $R_{\rm sh}$ for representative limiting cases of the parameters $\ell$, $\eta$, and $Q$. The table shows how Lorentz violation, electric charge, and the conical deficit modify the black hole's optical scales.}
\end{table*}

Fig.~\ref{fig:ShadowPara} provides a parametric plot illustrating how the angular radius of the black hole shadow varies with different values of the parameters $\ell$ and $\eta$. The results indicate that the shadow size tends to increase as either parameter increases. Physically, this suggests that both the Lorentz-violating parameter $\ell$ and the global monopole parameter $\eta$ act to diminish the strength of the gravitational field, which manifests as an enlarged apparent size of the black hole shadow. Additionally, Table~1 presents a summary of the photon sphere radius $r_s$ and the shadow radius $R_{\rm sh}$ for various limiting cases involving the parameters $\ell$, $\eta$, and $Q$.

\begin{figure*}[tbhp]
	\centerline{
		\includegraphics[width=180mm,height=140mm]{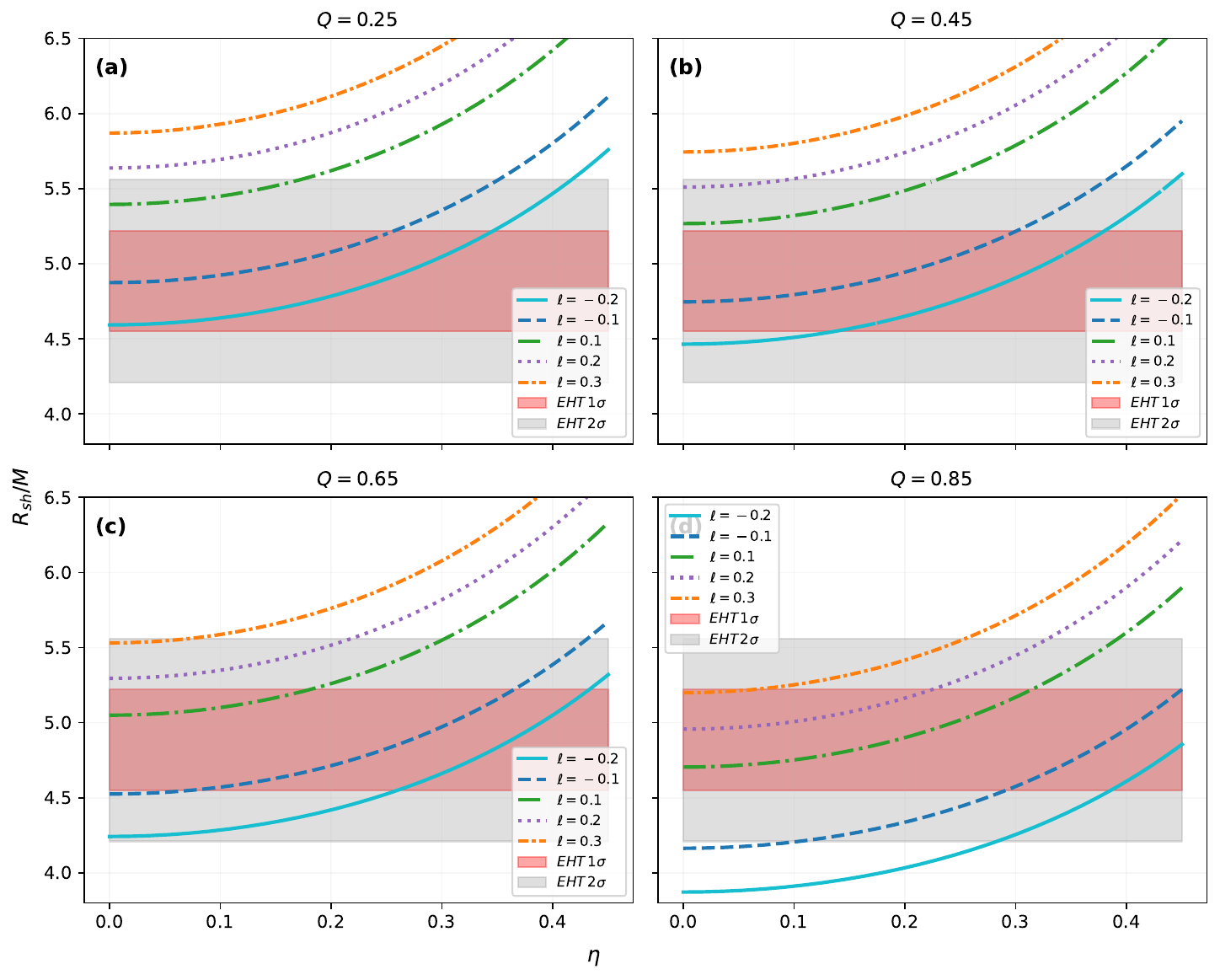}}
	\caption{Normalized shadow radius $R_{\rm sh}/M$ as a function of the monopole parameter $\eta$ for selected values of the charge $Q$ and the Lorentz-violating parameter $\ell$. The yellow and gray shaded bands represent the $1\sigma$ and $2\sigma$ EHT constraints from Sgr~A*, respectively. The intersections between the model curves and these bands indicate the observationally allowed parameter ranges.}
\label{fig:ShadowEHT}
\end{figure*}
\noindent Furthermore, we constrain the parameters $\ell$ and $\eta$ using the EHT observational results for Sgr A*. The numerical bounds adopted from these observations are summarized in Table~\ref{shadow_bounds}. The adopted methodology is widely used in the existing literature, and some recent studies are cited herein~\cite{Jiang:2023img,Ali:2024ssf,Tan:2025vrp,Kala:2025xnb,Kala:2025fld,Kala:2026bmz}.
\begin{table}[h]
\centering
\caption{Constraints on the shadow radius of Sgr A* obtained from EHT observations using Keck and VLTI mass-to-distance measurements~\cite{EHTL12}.}
\begin{tabular}{c c}
\hline
\textbf{Confidence Level} & $\mathbf{R_{\rm sh}/M}$ \\
\hline
$1\sigma$ bound & $4.55 \le R_{\rm sh}/M \le 5.22$ \\
$2\sigma$ bound & $4.21 \le R_{\rm sh}/M \le 5.56$ \\
\hline
\end{tabular}
\label{shadow_bounds}
\end{table}
\begin{table}[h!]
\centering
\caption{Estimated bounds on the parameters $\ell$ and $\eta$ for different values of the charge parameter $Q$, inferred from consistency with the $1\sigma$ and $2\sigma$ EHT intervals shown in Fig.~\ref{fig:ShadowEHT}.}
\begin{tabular}{c c c c}
\hline
$Q$ & Confidence Level & Allowed $\ell$ & Allowed $\eta$ \\
\hline
0.25 & $1\sigma$ & $\ell \lesssim 0.2$ & $\eta \lesssim 0.30$ \\
     & $2\sigma$ & $\ell \lesssim 0.3$ & $\eta \lesssim 0.40$ \\
\hline
0.45 & $1\sigma$ & $\ell \lesssim 0.15$ & $\eta \lesssim 0.28$ \\
     & $2\sigma$ & $\ell \lesssim 0.25$ & $\eta \lesssim 0.38$ \\
\hline
0.65 & $1\sigma$ & $\ell \lesssim 0.10$ & $\eta \lesssim 0.25$ \\
     & $2\sigma$ & $\ell \lesssim 0.20$ & $\eta \lesssim 0.35$ \\
\hline
0.85 & $1\sigma$ & $\ell \lesssim 0.05$ & $\eta \lesssim 0.20$ \\
     & $2\sigma$ & $\ell \lesssim 0.15$ & $\eta \lesssim 0.30$ \\
\hline
\end{tabular}
\label{tab:bounds}
\end{table}

Fig.~\ref{fig:ShadowEHT} depicts how the normalized shadow radius $R_{\rm sh}/M$ varies with the parameter $\eta$ for different values of the charge parameter $Q$. Each panel corresponds to a specific $Q$, while the colored curves indicate various choices of the Lorentz-violating parameter $\ell$. For a fixed $Q$, the shadow radius shows a monotonic increase as $\eta$ grows. Likewise, at a given $\eta$, larger $\ell$ values correspond to an enlarged shadow size. Physically, this pattern suggests that both the Lorentz-violating parameter $\ell$ and the global monopole parameter $\eta$ tend to weaken the gravitational field. This effect allows photons to maintain orbits at greater radii, which in turn leads to a larger apparent black hole shadow. The yellow and gray shaded areas represent the $1\sigma$ and $2\sigma$ confidence intervals derived from the EHT VLBI observations of Sgr~A$^*$, respectively.

By comparing the theoretical predictions with these observational constraints, it becomes possible to restrict the range of model parameters. Specifically, only those pairs of $\ell$ and $\eta$ that result in a shadow radius falling within the $1\sigma$ and $2\sigma$ bounds align with current observations. The figure indicates that moderate values of $\ell$, approximately $\ell \lesssim 0.2$--$0.3$, together with $\eta$ values up to about $\eta \lesssim 0.4$, generally remain compatible with the EHT data. Higher values of either parameter tend to produce shadow sizes exceeding the permissible ranges. Additionally, increasing the charge parameter $Q$ narrows these bounds, implying that larger charges require $\ell$ and $\eta$ to be smaller in order to maintain agreement with the measurements.

\section{Photon Trajectories in Weak Gravitational Field}

In this section, we study photon trajectories and analyze how the LV parameter and the global monopole affect photon paths.

To study the propagation of photons, we consider null geodesics ($ds^2=0$) restricted to the equatorial plane $(\theta=\pi/2)$. The corresponding optical metric can be written as~\cite{Huang:2023bto,kala2025propagation}
\begin{equation}
dl^{2}=\alpha_{rr}(r)\,dr^{2}+\alpha_{\phi\phi}(r)\,d\phi^{2},
\end{equation}
where
\begin{equation}
\alpha_{rr}(r)=\frac{1+\ell}{f(r)^2},
\qquad
\alpha_{\phi\phi}(r)=\frac{\beta^2 r^{2}}{f(r)}.
\end{equation}
Following the generalized Gauss--Bonnet method, the function associated with the Gaussian curvature of the optical manifold is defined as~\cite{Huang:2023bto}
\begin{equation}
H(r)=-\frac{\partial_r \alpha_{\phi\phi}}{2\sqrt{\alpha}},
\end{equation}
where $\alpha=\alpha_{rr}\alpha_{\phi\phi}$ denotes the determinant of the optical metric.

Expanding in the weak-field limit and evaluating along the photon trajectory $r_\gamma$, we obtain
\begin{equation}
H(r_\gamma)=
-\frac{1}{\sqrt{1+\ell}}
+\frac{2M\zeta}{r_\gamma\sqrt{1+\ell}}
-\frac{\varsigma Q^2}{r_\gamma^2\sqrt{1+\ell}}
+\mathcal{O}(M^2,Q^4).
\end{equation}

Introducing the inverse radial coordinate $u=1/r$, the null geodesic equation becomes
\begin{equation}
\left(\frac{du}{d\phi}\right)^2=\frac{\beta^2}{(1+\ell)b^2}-\frac{\beta^2u^2}{1+\ell}+\frac{2M\zeta\beta^2}{1+\ell}u^3-\frac{\varsigma \beta^2 Q^2}{1+\ell}u^4.
\end{equation}
where $b$ denotes the impact parameter.

It is convenient to introduce the rescaled angular variable
\[
x=\frac{\beta\phi}{\sqrt{1+\ell}}.
\]
In terms of $x$, the orbit equation takes the form
\[
\frac{d^2u}{dx^2}+u=3M\zeta\,u^2-2\varsigma Q^2\,u^3.
\]

Solving perturbatively in the weak-field regime yields the photon trajectory
\begin{equation}
\begin{split}
u(\phi)=&
\frac{\beta}{b}
\sin\!\left(\frac{\beta\phi}{\sqrt{1+\ell}}\right) 
+\frac{M\zeta\beta^2}{b^2}
\left[1+\cos^2\!\left(\frac{\beta\phi}{\sqrt{1+\ell}}\right)\right] \\
&-\frac{\varsigma Q^2\beta^3}{2b^3}
\sin^3\!\left(\frac{\beta\phi}{\sqrt{1+\ell}}\right)
+\mathcal{O}(M^2,Q^4).
\end{split}
\end{equation}

At leading order, the angular coordinates of the receiver and source can be obtained from
\begin{equation}
\phi(u)=
\frac{\sqrt{1+\ell}}{\beta}\,
\arcsin\!\left(\frac{bu}{\beta}\right)
+\mathcal{O}(M,Q^2).
\end{equation}

Thus, the angular positions of the receiver and source are
\begin{equation}
\phi_R=\phi(u_R),
\qquad
\phi_S=\pi-\phi(u_S),
\end{equation}
where $u_R=1/r_R$ and $u_S=1/r_S$ denote the inverse radial distances.

Using the finite-distance Gauss--Bonnet deflection formula~\cite{Huang:2023bto,Pantig:2024lpg}
\begin{equation}
\hat{\alpha}=\int_{\phi_S}^{\phi_R}[1+H(r_\gamma)]\,d\phi,
\end{equation}
and taking the asymptotic limit in which the source and observer are located at infinity, $(u_S,u_R\rightarrow0)$, the weak deflection angle becomes
\begin{align}
\hat{\alpha}_{\infty}
&=
\pi\left(\frac{\sqrt{1+\ell}}{\beta}-1\right)
+\frac{4M\zeta\sqrt{1+\ell}}{b}
-\frac{3\pi\varsigma Q^2\,\beta\sqrt{1+\ell}}{4b^2}
\notag\\&+\mathcal{O}(M^2,Q^4).
\end{align}

The first term represents the topological contribution associated with the asymptotically conical geometry, while the remaining terms describe the mass and charge corrections to the weak deflection angle. This makes the weak-field lensing analysis particularly interesting in the present spacetime, since even the asymptotic geometry already contributes nontrivially to the bending of light.

\begin{figure*}[tbhp]
	\centerline{
		\includegraphics[width=80mm,height=70mm]{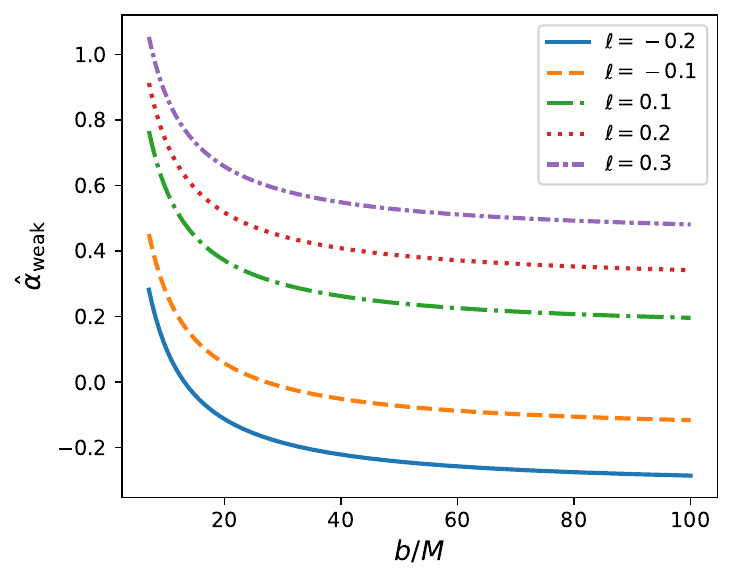}\qquad
        \includegraphics[width=80mm,height=70mm]{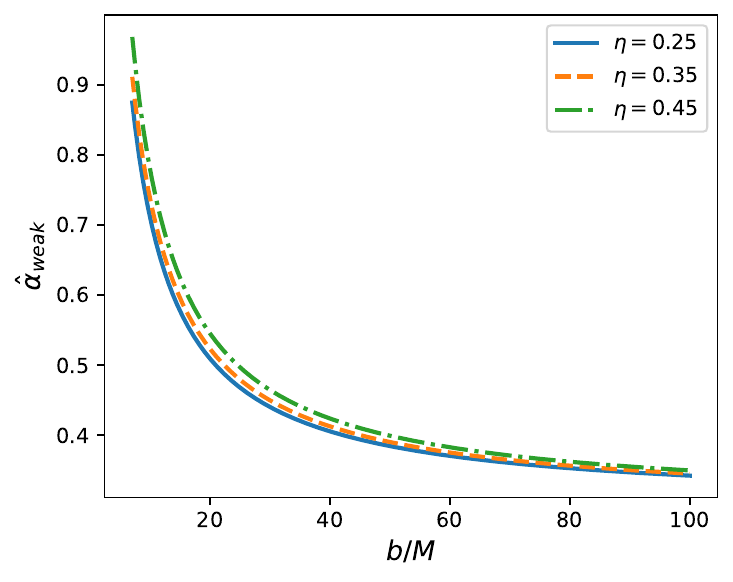}}(a) $\eta=0.3$ \hspace{5cm} (b) $\ell=0.2$
	\caption{Weak deflection angle $\hat{\alpha}_{\infty}$ as a function of the impact parameter $b/M$ for fixed $Q=0.8$. Panel (a) shows varying $\ell$ at fixed $\eta=0.3$, while panel (b) shows varying $\eta$ at fixed $\ell=0.2$. The curves include the topological contribution associated with the asymptotically conical geometry.}
\label{fig:weakdeflection}
\end{figure*}

From this expression, it is evident that the bending of light is governed by the effective parameters $\zeta$ and $\varsigma$, which incorporate both the Lorentz-violating parameter $\ell$ and the global monopole parameter $\eta$. The global monopole induces a solid angle deficit through the factor $\beta$, while the bumblebee parameter modifies the gravitational coupling appearing in the deflection angle.\\
Fig.~\ref{fig:weakdeflection} presents the variation of the weak deflection angle $\alpha_{\rm weak}$ as a function of the impact parameter $b/M$, examining different values of the parameters $\ell$ and $\eta$. The left panel focuses on the influence of $\ell$ while holding $\eta$ fixed at $0.3$, whereas the right panel explores changes in $\eta$ with $\ell$ held constant at $0.2$. In both cases, the deflection angle decreases steadily as the impact parameter grows, reflecting the typical behavior expected in the weak-field regime where light rays passing farther from the black hole experience diminished gravitational bending. The left panel reveals that increasing the Lorentz-violating parameter $\ell$ generally raises the deflection angle for a given value of $b/M$, indicating an enhancement of light bending induced by $\ell$ in this regime. On the other hand, the right panel shows that augmenting the global monopole parameter $\eta$ also leads to a larger deflection angle, though this effect appears less pronounced compared to the influence of $\ell$.

These observations imply that both parameters, $\ell$ and $\eta$, alter the spacetime geometry in a manner that intensifies gravitational lensing within the weak deflection limit. Nevertheless, the pronounced decline of $\alpha_{\rm weak}$ with increasing $b/M$ suggests that the impact of these parameters diminishes at greater distances. Since the spacetime contains a global monopole, the asymptotic geometry is conical rather than asymptotically flat.

\section{Solar System Test}

The motion of a test particle of mass $m$ along its geodesics can be described by the Lagrangian
\begin{equation}
\mathbb{L} = \frac{1}{2} m g_{\mu\nu} \frac{dx^{\mu}}{d\tau} \frac{dx^{\nu}}{d\tau}. \label{ss1}
\end{equation}
From the normalization condition for the four-velocity of timelike particles, one can write the Lagrangian as \(\mathbb{L}=\frac{\epsilon}{2}\), where \(\epsilon=-1\) for massive particles.

Since the spacetime is static and spherically symmetric, the Lagrangian is independent of $t$ and $\phi$. As a result, there are two conserved quantities: the energy $\mathcal{E}$ and the angular momentum $\mathcal{L}$ per unit mass of test particles, given by
\begin{align}
\mathcal{E}=f(r)\, \frac{dt}{d\tau},\qquad
\mathcal{L}= \beta\,r^2\, \frac{d\phi}{d\tau}.\label{ss2}
\end{align}

Then, from the conserved quantities in Eq.~(\ref{ss2}), for timelike geodesics, one can obtain a single differential equation for the radial coordinate $r$ in terms of the proper time $\tau$:
\begin{equation}
(1+\ell)\,\left(\frac{dr}{d\tau}\right)^2+\left(1+\frac{\mathcal{L}^2}{r^2}\right)\,f(r)= \mathcal{E}^2.\label{ss3}
\end{equation}

We now introduce the variable $v(\phi)=\frac{1}{r(\phi)}$, so that
\begin{equation}
\frac{dr}{d\tau}= \frac{dr}{d\phi}\frac{d\phi}{d\tau}= - \frac{\mathcal{L}}{\beta}\,\frac{dv}{d\phi}. \label{ss4}
\end{equation}

By substituting this into Eq.~(\ref{ss3}), we obtain
\begin{align}
(1+\ell) \left(\frac{dv}{d\phi}\right)^2 &+ \left(\beta^2+\varsigma\,\frac{\beta^2 Q^2}{\mathcal{L}^2}\right) v^2
=2 M \zeta \beta^2 v^3+\frac{\beta^2 (\mathcal{E}^2 - 1)}{\mathcal{L}^2}\nonumber\\
&\quad + \frac{2 M \zeta \beta^2}{\mathcal{L}^2} v -\varsigma \beta^2 Q^2 v^4. \label{ss5}
\end{align}

As is usually done in this treatment, it is preferable to solve the second-order differential equation obtained by differentiating the above equation with respect to $\phi$:
\begin{equation}
(1+\ell) \frac{d^2v}{d\phi^2}+\left(\beta^2+\varsigma\frac{\beta^2 Q^2}{\mathcal{L}^2}\right) v= \frac{M \zeta \beta^2}{\mathcal{L}^2}+ 3 M \zeta \beta^2 v^2-2\varsigma \beta^2 Q^2 v^3. \label{ss6}
\end{equation}

In order to solve Eq.~(\ref{ss6}), we set the electric charge $Q=0$ for simplicity. In this limiting case, the spacetime under consideration reduces to an uncharged Schwarzschild-like black hole with a global monopole. Therefore, Eq.~(\ref{ss6}) reduces to
\begin{equation}
(1+\ell) \frac{d^2v}{d\phi^2} +\beta^2 v = \frac{M \zeta \beta^2}{\mathcal{L}^2} + 3 M \zeta \beta^2 v^2. \label{ss7}
\end{equation}

Here, the Lorentz-violating parameter $\ell$ and the global monopole parameter $\eta$ enter through both the kinetic term and the coefficients of the effective force terms.

To solve Eq.~(\ref{ss7}) perturbatively, and noting that $\ell \ll 1$, it is valid to treat the last term as a relativistic correction compared with the Newtonian case. The perturbative solution is expressed in terms of a small parameter
\begin{equation}
v \simeq v^{(0)} + \epsilon v^{(1)}, \qquad \epsilon = \frac{3M^2 \zeta^2 \beta^2}{\mathcal{L}^2}, \label{ss8}
\end{equation}
where $v^{(0)}$ satisfies the zeroth-order differential equation
\begin{equation}
(1+\ell) \frac{d^2v^{(0)}}{d\phi^2} + \beta^2 v^{(0)} - \frac{M \zeta \beta^2}{\mathcal{L}^2} = 0. \label{ss9}
\end{equation}

The solution of Eq.~(\ref{ss9}) is
\begin{equation}
v^{(0)} = \frac{M \zeta \beta^2}{\mathcal{L}^2} \left[ 1 + e \, \cos \left(\frac{\beta \phi}{\sqrt{1+\ell}}\right) \right], \label{ss10}
\end{equation}
which is analogous to the Newtonian result. The integration constants are chosen as the orbital eccentricity $e$ (assumed small, similar to GR) and the initial condition $\phi_0 = 0$.

At first order in $\epsilon$, the differential equation reads
\begin{equation}
(1+\ell) \frac{d^2v^{(1)}}{d\phi^2} + \beta^2 v^{(1)} - \frac{\mathcal{L}^2}{M \zeta} \left(v^{(0)}\right)^2 = 0, \label{ss11}
\end{equation}
which admits an approximate solution of the form
\begin{align}
v^{(1)} &\simeq \frac{M \zeta \beta^2}{\mathcal{L}^2}\, \frac{e \beta \phi}{\sqrt{1+\ell}} \, \sin \left( \frac{\beta \phi}{\sqrt{1+\ell}} \right)\nonumber\\
&\quad+ \frac{M \zeta \beta^2}{\mathcal{L}^2} \left[ 1+\frac{e^2}{2} - \frac{e^2}{6} \cos \left(\frac{2 \beta \phi}{\sqrt{1+\ell}} \right) \right]. \label{ss12}
\end{align}

For practical purposes, the second term in Eq.~(\ref{ss12}) can be neglected, as it consists of a constant shift and an oscillatory contribution that averages to zero.

Therefore, the perturbative solution (\ref{ss7}) reads
\begin{align}
    v \simeq \frac{M \zeta \beta^2}{\mathcal{L}^2}  \Bigg[1+ e \cos \left( \frac{\beta \phi}{\sqrt{1+\ell}} \right)+\epsilon \frac{e \beta \phi}{\sqrt{1+\ell}} \sin \left( \frac{\beta \phi}{\sqrt{1+\ell}} \right) \Bigg].\label{ss13}
\end{align}

Because \(\epsilon \ll 1\), the perturbative solution (\ref{ss13}) can be rewritten in the form of an ellipse equation,
\begin{equation}
    v \simeq \frac{M \zeta \beta^2}{\mathcal{L}^2} \left[1+ e \cos \left( \frac{ \beta(1-\epsilon) \phi}{\sqrt{1+\ell}} \right)\right].\label{ss14} 
\end{equation}

Despite the presence of Lorentz violation and global monopole, the orbit remains periodic with a period $\Phi$,  
\begin{equation}
\Phi = 2 \pi \frac{\sqrt{1+\ell}}{\beta(1-\epsilon)} \approx 2 \pi + \Delta \Phi, \label{ss15}
\end{equation}
where $\Delta \Phi$ represents the advance of the perihelion.  

By expanding to first order in the small parameters $\ell$, $\eta^2$, and $\epsilon$, the perihelion shift $\Delta \Phi$ for the uncharged black hole is given by
\begin{align}
\Delta \Phi 
&= 2 \pi \left(\frac{\sqrt{1+\ell}}{\beta(1-\epsilon)}-1\right)
\simeq 2 \pi \epsilon + \pi \ell + \pi \eta^2
\notag\\
& \simeq \frac{6 \pi M^2}{\mathcal{L}^2}+\left(\frac{6 M}{\mathcal{L}^2}+1\right)\,\left(\ell + \eta^2\right) \pi.
\label{ss16}
\end{align}

Equation~(\ref{ss16}) shows that both the Lorentz-violating and global monopole parameters contribute explicitly to the perihelion shift in the weak-field approximation.

In the limit where $\ell = 0$ (absence of Lorentz-violating effects) and $\eta = 0$ (absence of global monopole effects), the perihelion shift $\Delta \Phi$ reduces to
\begin{equation}
    \Delta \Phi = 2 \pi \left(\frac{1}{1-\epsilon}-1\right) \simeq \frac{6 \pi M^2}{\mathcal{L}^2},
\end{equation}
which is the standard Schwarzschild result.

\begin{figure*}[tbhp]
	\centerline{
		\includegraphics[width=80mm,height=70mm]{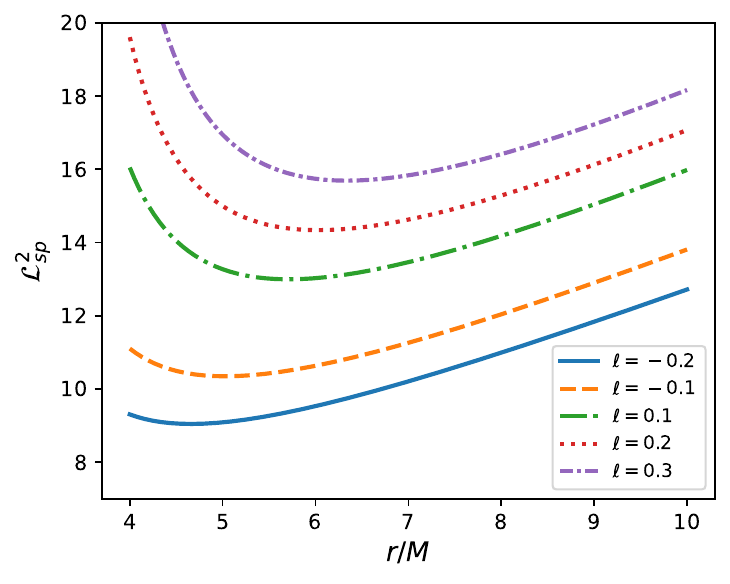}\qquad
        \includegraphics[width=80mm,height=70mm]{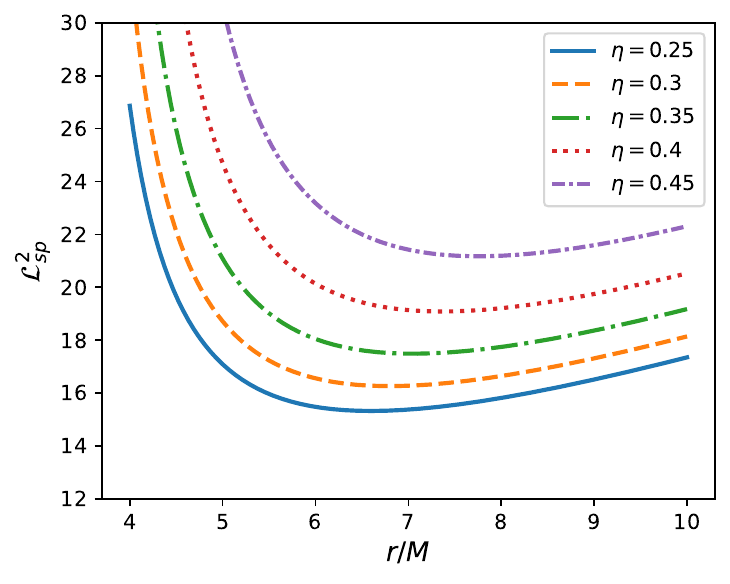}}(a) $\eta=0.3$ \hspace{5cm} (b) $\ell=0.2$
	\caption{Specific angular momentum squared, $L^{2}$, as a function of the radial coordinate for circular timelike motion. Panel (a) shows varying $\ell$ at fixed $\eta=0.3$, while panel (b) shows varying $\eta$ at fixed $\ell=0.2$. The minima of the curves are associated with the onset of marginal stability and hence with the ISCO trend.}
\label{fig:dynamicAngular}
\end{figure*}

\begin{figure*}[tbhp]
	\centerline{
		\includegraphics[width=80mm,height=70mm]{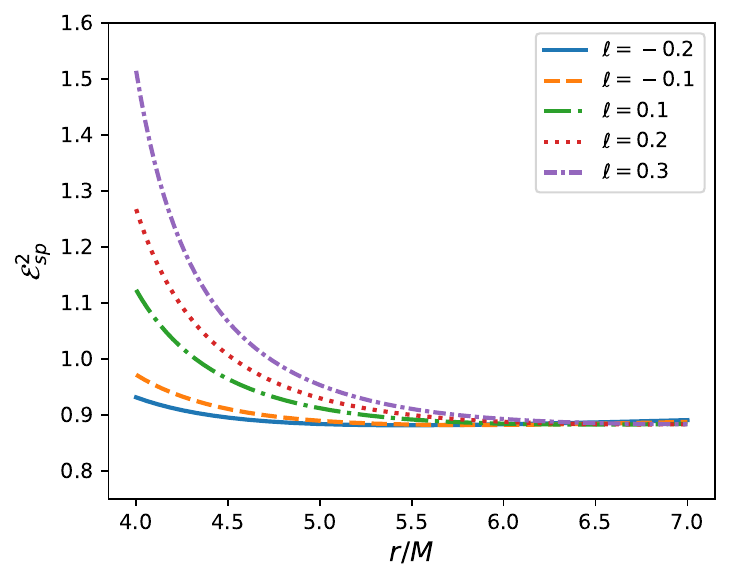}\qquad
        \includegraphics[width=80mm,height=70mm]{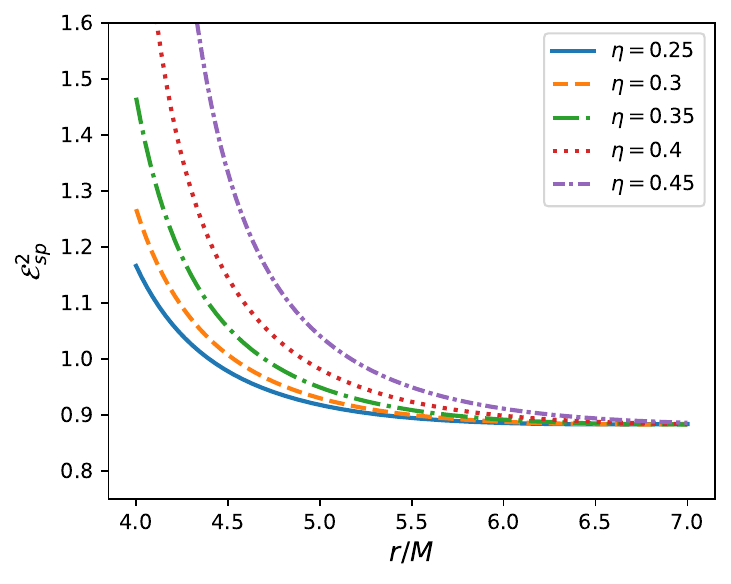}}(a) $\eta=0.3$ \hspace{5cm} (b) $\ell=0.2$
	\caption{Specific energy squared, $E^{2}$, as a function of the radial coordinate for circular timelike motion. Panel (a) corresponds to $\eta=0.3$ with varying $\ell$, and panel (b) corresponds to $\ell=0.2$ with varying $\eta$. The profiles indicate how much orbital energy is required to sustain bound motion at a given radius.}
\label{fig:dynamicEnergy}
\end{figure*}

\section{Dynamics of neutral Test Particles}

In this section, we investigate the dynamics of neutral test particles in the gravitational field of the LV black hole under consideration with a global monopole. Our analysis focuses on circular motion, specifically the characteristics of the innermost stable circular orbit (ISCO), given its fundamental role in accretion-disk physics and astrophysical phenomenology.

\begin{figure*}[tbhp]
	\centerline{
		\includegraphics[width=80mm,height=70mm]{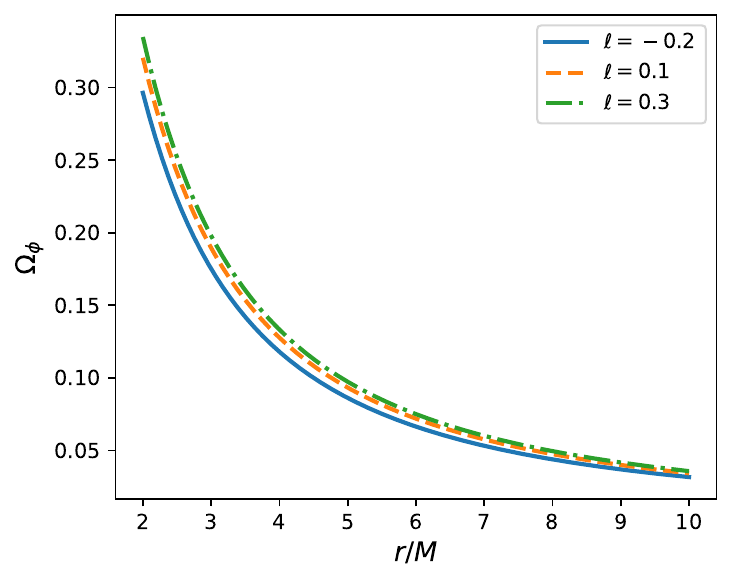}\qquad
        \includegraphics[width=80mm,height=70mm]{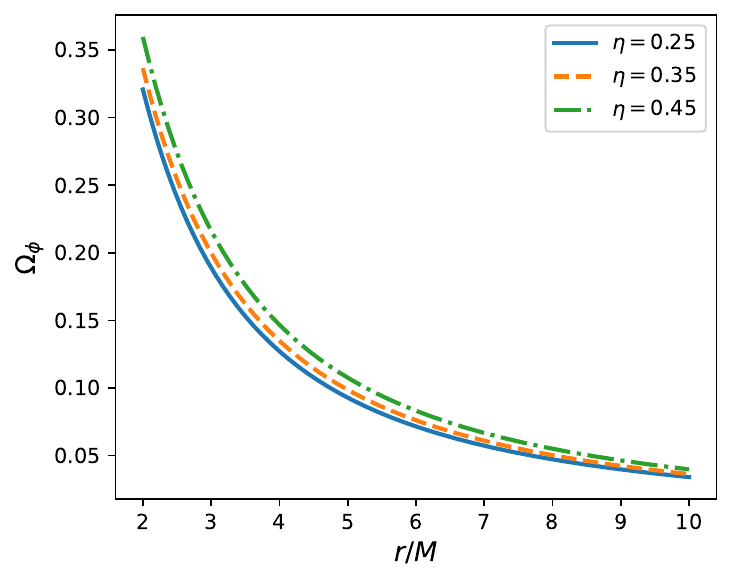}}(a) $\eta=0.3$ \hspace{5cm} (b) $\ell=0.2$
	\caption{Azimuthal angular velocity $\Omega_{\phi}$ as a function of the radial coordinate for neutral circular orbits. Panel (a) shows varying $\ell$ at fixed $\eta=0.3$, while panel (b) shows varying $\eta$ at fixed $\ell=0.2$. Larger values of the geometric parameters reduce the orbital frequency at a fixed radius.}
\label{fig:dynamicOmega}
\end{figure*}

The motion of a neutral test particle of mass $\mu$ along its geodesics can be described by the Lagrangian
\begin{equation}
\mathbb{L} = \frac{1}{2} \mu g_{\mu\nu} \frac{dx^{\mu}}{d\tau} \frac{dx^{\nu}}{d\tau}. \label{kk1}
\end{equation}
Using the conserved quantities \(\mathcal{E}=f(r)\,dt/d\tau\) and \(\mathcal{L}=\beta r^2\,d\phi/d\tau\), one can derive the corresponding radial equation of motion for neutral test particles.

Following the same steps as before, we obtain the equation of motion for the radial coordinate \(r\) as
\begin{equation}
(1+\ell)\,\left(\frac{dr}{d\tau}\right)^2=U_{\rm eff},\label{kk2}
\end{equation}
where the effective potential of the system is given by
\begin{equation}
    U_{\rm eff}=\mathcal{E}^2-\left(1+\frac{\mathcal{L}^2}{r^2}\right)\,f(r).\label{kk3}
\end{equation}

For circular orbits of fixed radii, the following conditions must be satisfied
\begin{equation}
U_{\rm eff}=0\Longrightarrow \mathcal{E}^2=\left(1+\frac{\mathcal{L}^2}{r^2}\right)\,f(r).\label{kk4}
\end{equation}
and
\begin{equation}
   \frac{\partial U_{\rm eff}}{\partial r}=0\Longrightarrow \frac{\partial }{\partial r}\left(f(r)+\frac{\mathcal{L}^2}{r^2}\,f(r)\right)=0.\label{kk5}
\end{equation}

Simplification of these relations yields
\begin{align}
\mathcal{L}^2_{\rm sp}&=\frac{r ^3\,f'(r)}{2 f(r)-r\,f'(r)},\label{kk6}\\
\mathcal{E}^2_{\rm sp}&=\frac{2 f^2 (r)}{2 f(r)-r\,f'(r)}.\label{kk7}
\end{align}

Here $\mathcal{L}_{\rm sp}$ and $\mathcal{E}_{\rm sp}$, respectively, are the specific angular momentum and specific energy of test particles.

Moreover, the orbital velocity of the test particles revolving in circular orbits of fixed radii along the azimuthal direction is given by
\begin{equation}
    \Omega_{\phi}\Big{|}_{r=\mbox{const.}}=\frac{d\phi}{dt}\Big{|}_{r=\mbox{const.}}=\frac{\mathcal{L}_{\rm sp}}{\mathcal{E}_{\rm sp}}\,\frac{f(r)}{r^2}=\sqrt{\frac{f'(r)}{2r}},\label{kk8}
\end{equation}
where we have used Eq.~(\ref{kk7}).

In addition to satisfying the conditions for circular motion, the stability of such orbits requires that the effective potential $U_{\rm eff}$ must satisfy the following conditions:
\begin{equation}
U_{\rm eff}=0,\quad \frac{\partial U_{\rm eff}}{\partial r}=0,\quad \frac{\partial^2 U_{\rm eff}}{\partial r^2} \leq 0.\label{kk9}
\end{equation}

For marginally stable circular orbits, we have $\frac{\partial^2 U_{\rm eff}}{\partial r^2}=0$, which results
\begin{equation}
    f(r)\,f''(r)-2 (f'(r))^2+\frac{3 f(r)\,f'(r)}{r}=0.\label{kk10}
\end{equation}
Figs.~\ref{fig:dynamicAngular} to~\ref{fig:dynamicOmega} depict the behavior of key dynamical variables that govern the motion of neutral test particles within the given spacetime, analyzed as functions of the radial coordinate $r/M$ for varying values of the parameters $\ell$ and $\eta$, while keeping the charge fixed at $Q = 0.8$. Fig.~\ref{fig:dynamicAngular} illustrates how the specific angular momentum $L^{2}$ varies with radial distance. It shows an initial decrease, followed by a minimum, and then an increase as $r/M$ grows. This pattern suggests the presence of stable circular orbits. The location of the minimum shifts in response to changes in $\ell$ and $\eta$, indicating that these parameters influence the radius of the innermost stable circular orbit (ISCO). Typically, larger values of $\ell$ and $\eta$ correspond to greater angular-momentum requirements, reflecting changes in the effective gravitational potential felt by neutral particles. Fig.~\ref{fig:dynamicEnergy} presents the specific energy $E^{2}$ as a function of $r/M$. The energy decreases monotonically as the radial distance increases, approaching a constant value at large $r/M$, consistent with the asymptotically conical behavior of the spacetime. Rising values of $\ell$ and $\eta$ slightly elevate the energy levels, implying that particles require more energy to maintain bound orbits. This trend suggests a reduction in the effective gravitational attraction. Fig.~\ref{fig:dynamicOmega} shows the angular velocity $\Omega$ of neutral test particles plotted against $r/M$. As anticipated in Keplerian-like motion, $\Omega$ decreases monotonically with radius. At fixed $r/M$, higher $\ell$ and $\eta$ values correspond to reduced angular velocity, indicating slower orbital movement. This observation further reinforces the interpretation that both the LV and GM parameters effectively weaken the gravitational influence.

In summary, the parameters $\ell$ and $\eta$ notably affect the dynamics of neutral test particles by altering the effective potential, shifting the conditions for orbital stability, and diminishing the gravitational pull. These influences consistently emerge across the profiles of angular momentum, energy, and angular velocity.

\section{Scalar Perturbations}

The study of scalar perturbations of black holes provides a fundamental framework for investigating wave dynamics, linear stability, and quasinormal modes in curved spacetime \cite{Vishveshwara1970, Chandrasekhar1983, Ferrari1984, Kokkotas1999, Berti2009, Konoplya2011}. Numerous studies of scalar field perturbations have been conducted in various black hole configurations (see, \cite{FA7,FA8} and related references therein)

The massless scalar field wave equation is described by the Klein-Gordon equation as follows:
\begin{equation}
\frac{1}{\sqrt{-g}}\,\partial_{\mu}\left(\sqrt{-g}\,g^{\mu\nu}\,\partial_{\nu}\Psi\right)=0\quad\quad (\mu,\nu=0,\cdots,3), \label{mm1}    
\end{equation}
where $\Psi$ is the wave function of the scalar field, $g_{\mu\nu}$ is the covariant metric tensor, $g=\det(g_{\mu\nu})$ is the determinant of the metric tensor, $g^{\mu\nu}$ is the contravariant form of the metric tensor, and $\partial_{\mu}$ is the partial derivative for the coordinate systems.

Let us consider the following ansatz for the scalar field wave function:
\begin{equation}
    \Psi(t, r,\theta, \phi)=\exp(-i\,\omega\,t)\,Y^{\bar m}_{\bar \ell} (\theta,\phi)\,\frac{\psi(r)}{r},\label{mm2}
\end{equation}
where $\omega$ is the (possibly complex) temporal frequency, $\psi(r)$ is the radial part of the scalar field, and $Y^{\bar m}_{\bar \ell}(\theta,\phi)$ is the spherical harmonic with $\{\bar m,\, \bar \ell\}$ are the magnetic and orbital quantum numbers related by ${\bar \ell} \geq |{\bar m}|$.

Explicitly writing the wave equation (\ref{mm1}) using the metric (\ref{metric}) and then Eq.~(\ref{mm2}), we find
\begin{equation}
    \frac{\partial^2 \psi(r_*)}{\partial r_*^2}+\left(\omega^2-V_0\right)\,\psi(r_*)=0,\label{mm3}
\end{equation}
where the coordinate transformation (called the tortoise coordinate, $r_*$) is
\begin{eqnarray}
    r_*=\int\,\frac{\sqrt{1+\ell}\,dr}{f(r)}\quad,\quad \partial_{r_*}=\frac{f(r)}{\sqrt{1+\ell}}\,\partial_r.\label{mm4}
\end{eqnarray}
We have also used the identity
\begin{equation}
\left[ \frac{1}{\sin\theta} \frac{\partial}{\partial\theta} \left( \sin\theta \frac{\partial}{\partial\theta} \right) + \frac{1}{\sin^2\theta} \frac{\partial^2}{\partial\phi^2} \right] Y^{\bar m}_{\bar \ell}(\theta, \phi) = -{\bar \ell}\,({\bar \ell}+1) Y^{\bar m}_{\bar \ell} (\theta, \phi),\label{mm6} 
\end{equation}
with $Y^{\bar m}_{\bar \ell} (\theta, \phi)=e^{i\,{\bar m}\,\phi} P^{\bar \ell}_{\bar m}(\cos \theta)$.

The scalar perturbation potential $V_0(r)$ is given by
\begin{equation}
V_0(r)=\left(\frac{{\bar \ell}\,({\bar \ell}+1)}{\beta^2 r^2}+\frac{f'(r)}{r}\right)\,f(r).\label{mm5}
\end{equation}

\begin{figure*}[tbhp]
	\centerline{
		\includegraphics[width=80mm,height=70mm]{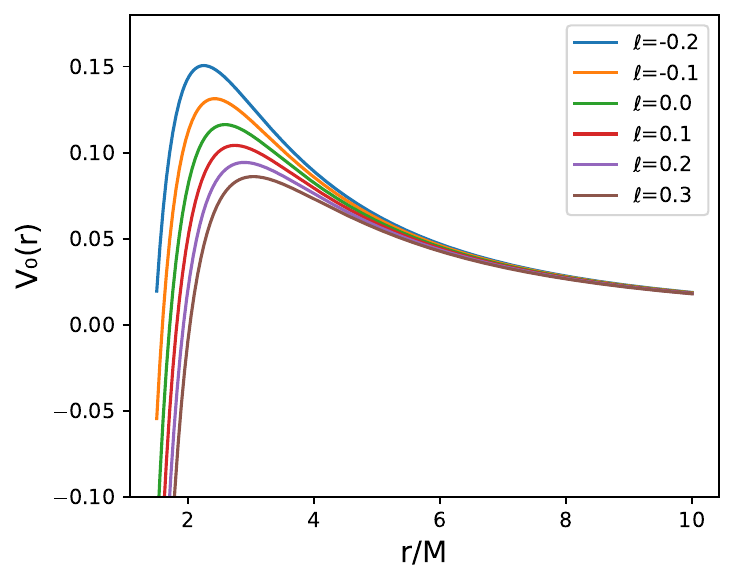}\qquad
        \includegraphics[width=80mm,height=70mm]{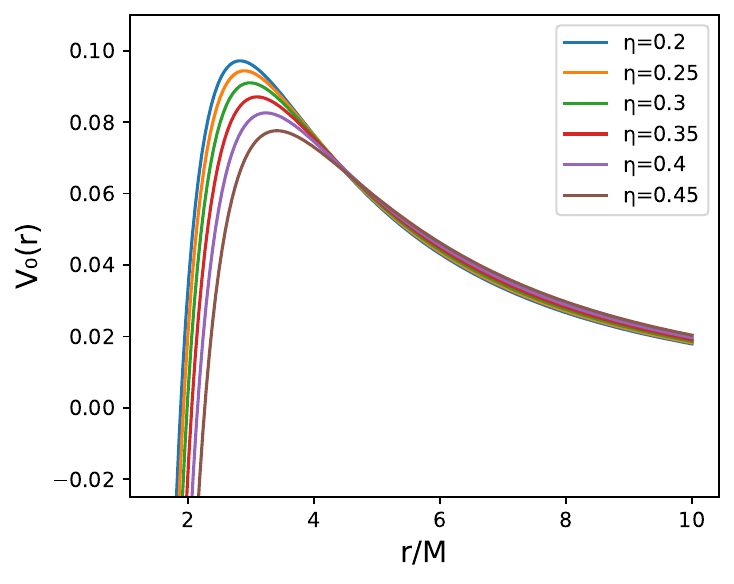}}(a) $\eta=0.3$ \hspace{5cm} (b) $\ell=0.2$
	\caption{Scalar potential as a function of the radial distance $r$ for fixed $\bar \ell=1$ and $Q=0.8$. Panel (a) corresponds to $\eta=0.25$ with varying $\ell$, while panel (b) corresponds to $\ell=0.2$ with varying $\eta$. }
\label{fig:scalarpotential}
\end{figure*}

We observe that the perturbation potential is influenced by the geometric parameters $\{\ell,\eta,M,Q\}$. Moreover, the multipole mode ${\bar \ell}$ also modifies this potential, so the scalar sector provides an additional probe of the spacetime beyond thermodynamics and null geodesics. Fig.~\ref{fig:scalarpotential} depicts the behavior of the scalar effective potential as a function of the radial coordinate $r$ for different values of the parameters $\ell$ and $\eta$, with fixed $\bar{\ell}=1$ and $Q=0.8$. The potential exhibits a single peak outside the event horizon, corresponding to the potential barrier that governs wave propagation. In panel (a), for fixed $\eta=0.25$, increasing $\ell$ modifies the height and location of the peak, indicating that the LV parameter influences the strength of the barrier. In panel (b), for fixed $\ell=0.2$, varying $\eta$ alters the potential profile, generally reducing the barrier height. Collectively, the parameters $\ell$ and $\eta$ play a crucial role in shaping the effective potential, which in turn governs the wave scattering behavior and the characteristics of the QNM spectrum of the black hole.

\subsection{Quasinormal modes}

Quasinormal modes (QNMs) describe the characteristic damped oscillations of a black hole under perturbations and are determined by imposing purely ingoing boundary conditions at the event horizon and purely outgoing conditions at spatial infinity~\cite{Konoplya2011,Kokkotas1999,Berti2009}. These modes are characterized by complex frequencies $\omega = \omega_R + i\,\omega_I$, where $\omega_R$ represents the oscillation frequency and $\omega_I$ determines the damping rate. Over the past few decades, QNMs have been extensively studied using various analytical and numerical techniques~\cite{Konoplya2011}. Some recent studies of QNMs in black holes can be found in~\cite{Pantig2025,Kala:2025kkm,Pantig2026} and related references therein.

To compute the QNMs, we employ the higher-order WKB approximation, which provides improved accuracy compared to the first-order treatment, especially for low-lying modes. In the 6th-order WKB approach, the QNM frequencies are determined by the relation \cite{Konoplya2011}
\begin{equation}
\frac{i\left(\omega^2 - V_0\right)}{\sqrt{-2V_0''}} - \sum_{j=2}^{6} \Lambda_j = n + \frac{1}{2},
\end{equation}
where $V_0$ is the maximum of the effective potential and $V_0''$ is its second derivative with respect to the tortoise coordinate evaluated at the peak $r=r_0$. The terms $\Lambda_j$ ($j=2,\dots,6$) represent higher-order correction terms that depend on higher derivatives of the potential at $r_0$. Solving for $\omega$, one obtains the complex quasinormal frequencies $\omega = \omega_R + i\,\omega_I$, where the real and imaginary parts are influenced not only by the peak value of the potential but also by its higher-order curvature properties. In this framework, $\omega_R$ is primarily determined by the height of the effective potential barrier, while $\omega_I$ depends on the curvature and higher-order derivatives, which encode the detailed shape of the potential near its maximum. The inclusion of higher-order correction terms $\Lambda_j$ significantly improves the accuracy of the WKB method, making it reliable for a wider range of parameters and modes.

\begin{figure*}[tbhp]
	\centerline{
		\includegraphics[width=170mm,height=80mm]{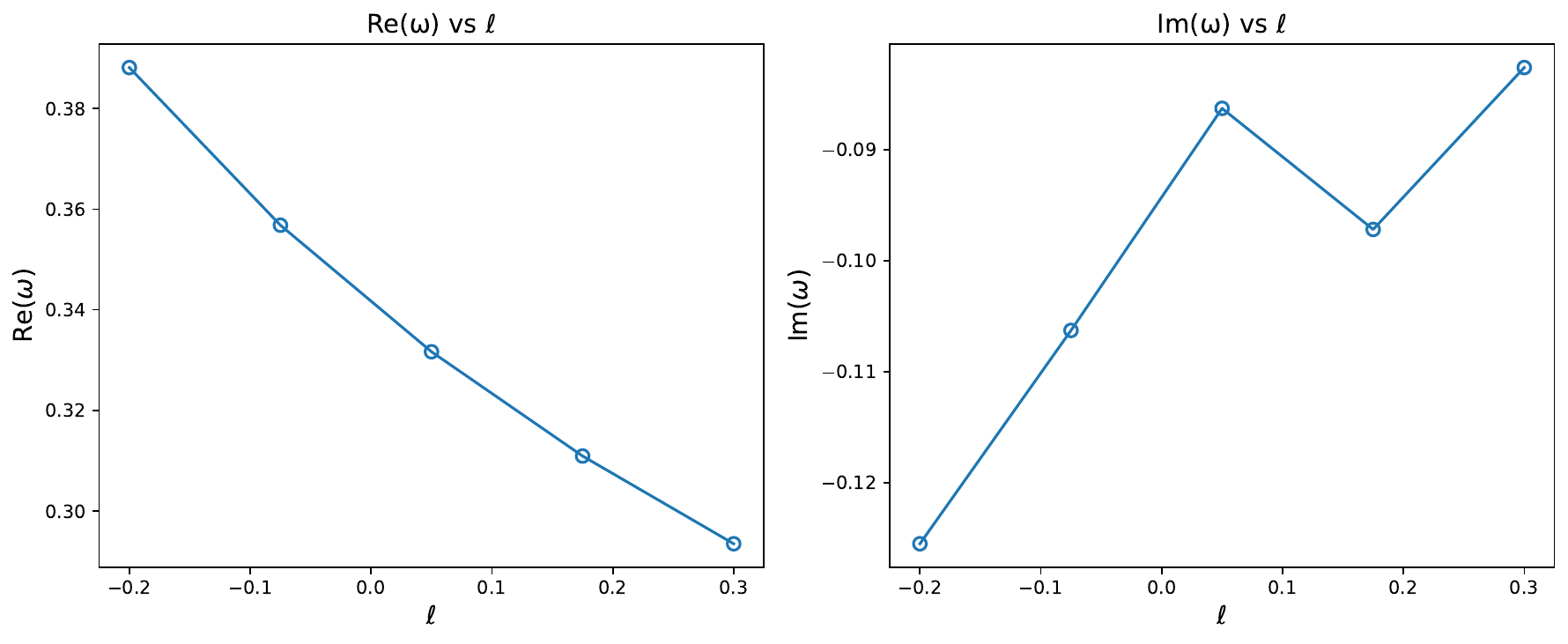}}
	\caption{Variation of the real and imaginary parts of the quasinormal mode frequency, Re$(\omega)$ and Im$(\omega)$, as functions of the LV parameter $\ell$ for a fixed GM parameter $\eta = 0.25$. The results correspond to scalar perturbations with $Q=0.8$, $n=0$, and $\bar{\ell}=1$.}
\label{fig:QNMs1}
\end{figure*}
Fig.~\ref{fig:QNMs1} shows the variation of the real and imaginary parts of the QNMs frequency, $\mathrm{Re}(\omega)$ and $\mathrm{Im}(\omega)$, as functions of the LV parameter $\ell$, while keeping the other parameters fixed. It is observed that the real part of the frequency decreases monotonically with increasing $\ell$, indicating that the oscillation frequency of the perturbations is suppressed in the presence of stronger Lorentz violation. On the other hand, the imaginary part exhibits a non-monotonic behavior. Initially, $\mathrm{Im}(\omega)$ becomes less negative, implying a slower decay of the perturbations. However, for larger values of $\ell$, the magnitude of $\mathrm{Im}(\omega)$ increases again, indicating a faster damping rate. This suggests that the LV parameter has a nontrivial influence on both the oscillatory and damping properties of the system.

\begin{figure*}[tbhp]
	\centerline{
		\includegraphics[width=170mm,height=80mm]{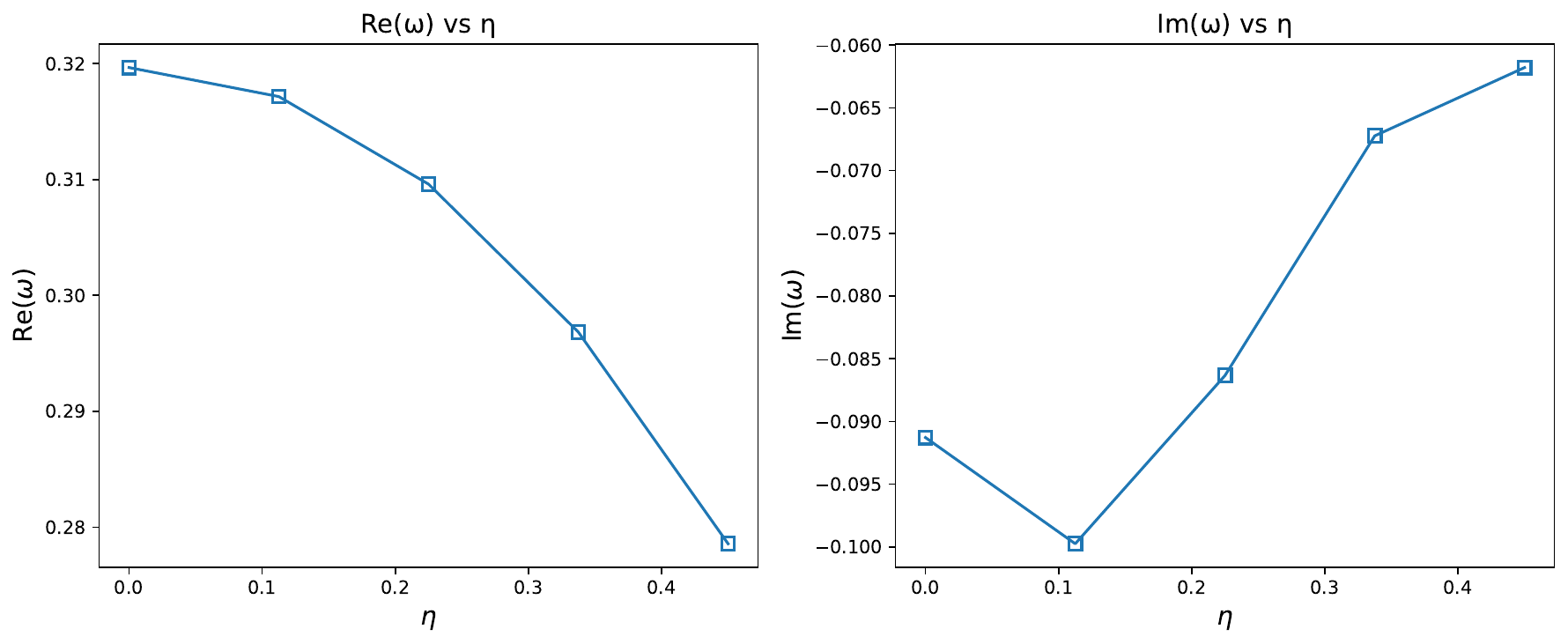}}
	\caption{Variation of the real and imaginary parts of the quasinormal mode frequency, Re$(\omega)$ and Im$(\omega)$, as functions of the GM parameter $\eta$ for a fixed LV parameter $\ell = 0.2$. The results correspond to scalar perturbations with $Q=0.8$, $n=0$, and $\bar{\ell}=1$.}
\label{fig:QNMs2}
\end{figure*}
Fig.~\ref{fig:QNMs2} illustrates the dependence of the real and imaginary parts of the QNMs frequency on the GM parameter $\eta$, with all other parameters held fixed. The real part of the frequency decreases steadily with increasing $\eta$, which indicates that the presence of a stronger monopole field reduces the oscillation frequency of the perturbations. In contrast, the imaginary part shows a non-monotonic trend, it first becomes more negative, corresponding to a faster decay of perturbations, and then increases (becomes less negative) for higher values of $\eta$, indicating a slower decay rate. This behavior demonstrates that the monopole parameter significantly influences the damping characteristics and overall stability of the perturbations.

\begin{table}
\centering
\caption{Quasinormal mode frequencies of charged black holes in bumblebee gravity with GM for scalar perturbations with $Q=0.8$, $n=0$, and $\bar{\ell}=1$.}
\begin{tabular}{|c|c|c|c|}
\hline
$\ell$ & $\eta$ & Re($\omega$) & Im($\omega$) \\
\hline
-0.20 & 0.00 & 0.403760 & -0.131869 \\
-0.20 & 0.15 & 0.398164 & -0.120275 \\
-0.20 & 0.30 & 0.381179 & -0.129966 \\
-0.20 & 0.45 & 0.352128 & -0.105682 \\

-0.08 & 0.00 & 0.371208 & -0.111442 \\
-0.08 & 0.15 & 0.366050 & -0.108431 \\
-0.08 & 0.30 & 0.350401 & -0.105966 \\
-0.08 & 0.45 & 0.323639 & -0.092073 \\

0.05 & 0.00 & 0.345073 & -0.097515 \\
0.05 & 0.15 & 0.340270 & -0.107456 \\
0.05 & 0.30 & 0.325696 & -0.097285 \\
0.05 & 0.45 & 0.300779 & -0.089503 \\

0.17 & 0.00 & 0.323533 & -0.096405 \\
0.17 & 0.15 & 0.319023 & -0.087439 \\
0.17 & 0.30 & 0.305339 & -0.084203 \\
0.17 & 0.45 & 0.281946 & -0.089397 \\

0.30 & 0.00 & 0.305411 & -0.077077 \\
0.30 & 0.15 & 0.301147 & -0.089422 \\
0.30 & 0.30 & 0.288214 & -0.091053 \\
0.30 & 0.45 & 0.266108 & -0.061240 \\
\hline
\end{tabular}
\label{tab:QNMresults}
\end{table}

\subsection{Greybody Factors}

The purpose of this section is to analyze the greybody factor (GF) for the considered black hole spacetime. The curvature of spacetime and the gravitational potential barrier surrounding a black hole modify the spectrum of the Hawking radiation it emits. As a result, the observed radiation spectrum deviates from that of an ideal black body. This modified spectrum is referred to as the greybody factor, since it incorporates the effects of the surrounding geometry, which acts as a potential barrier to the outgoing radiation. Consequently, the GF quantifies the deviation between the emitted Hawking radiation and the spectrum measured by an asymptotic observer \cite{Visser1999,Fernando2005,Boonserm2008}.

A number of theoretical studies have investigated greybody factors, or equivalently, the transmission probabilities of Hawking radiation, for various black hole solutions in different gravity models using diverse analytical and numerical techniques. In this work, we employ the general semi-analytic bounds approach to determine bounds on the black hole's greybody factor. The general bounds on the transmission and reflection coefficients are given by \cite{Boonserm2008,Konoplya2011,Boonserm2019,Kokkotas1999}
\begin{equation}
T(\omega) \geq \mbox{sech}^2 \left( \int_{-\infty}^{+\infty} \varrho\, dr_* \right),
\label{gf1}
\end{equation}
and
\begin{equation}
R(\omega) \leq \tanh^2 \left( \int_{-\infty}^{+\infty} \varrho\, dr_* \right),
\label{gf2}
\end{equation}
respectively.

Here, $\varrho$ is defined as
\begin{equation}
\varrho = \frac{\sqrt{\left(Y'\right)^2 + \left(\omega^2 - V_{0} - Y^2\right)^2}}{2Y},
\label{gf3}
\end{equation}
where $Y$ is a positive auxiliary function satisfying $Y(r_*) > 0$ and $Y(+\infty) = Y(-\infty) = \omega$.

To simplify the analysis, we choose
\begin{equation}
Y^2 = \omega^2,
\label{gf4}
\end{equation}
which significantly reduces the complexity of the expressions. With this choice, the bounds on the transmission and reflection coefficients reduce to
\begin{equation}
T(\omega) \geq \mbox{sech}^2 \left( \frac{1}{2 \omega} \int_{r_h}^{+\infty} V_0\, dr_* \right),
\label{gf5}
\end{equation}
and
\begin{equation}
R(\omega) \leq \tanh^2 \left( \frac{1}{2 \omega} \int_{r_h}^{+\infty} V_0\, dr_* \right),
\label{gf6}
\end{equation}
respectively. Here, $r_h$ is the event horizon of the black hole.

Using Eq.~(\ref{mm5}) together with the tortoise-coordinate relation (\ref{mm4}), one finds
\begin{equation}
\int_{r_h}^{+\infty} V_0\,dr_*
=
\sqrt{1+\ell}
\left[
\frac{{\bar \ell}\,({\bar \ell}+1)}{\beta^2 r_h}
+\frac{M\zeta}{r_h^2}
-\frac{2\varsigma Q^2}{3r_h^3}
\right].
\end{equation}
Therefore, the transmission coefficient satisfies
\begin{equation}
T(\omega)\ge
\mbox{sech}^2\left\{
\frac{\sqrt{1+\ell}}{2\omega}
\left[
\frac{{\bar \ell}\,({\bar \ell}+1)}{\beta^2 r_h}
+\frac{M\zeta}{r_h^2}
-\frac{2\varsigma Q^2}{3r_h^3}
\right]
\right\}.
\label{gf7}
\end{equation}
Similarly, the reflection coefficient satisfies
\begin{equation}
R(\omega)\le
\tanh^2\left\{
\frac{\sqrt{1+\ell}}{2\omega}
\left[
\frac{{\bar \ell}\,({\bar \ell}+1)}{\beta^2 r_h}
+\frac{M\zeta}{r_h^2}
-\frac{2\varsigma Q^2}{3r_h^3}
\right]
\right\}.
\label{gf8}
\end{equation}

Here $\zeta$ and $\varsigma$ are given in Eq.~(\ref{fucntion-3}). These bounds make explicit how the Lorentz-violating parameter, the monopole-induced deficit angle, and the electric charge jointly affect the transmission of Hawking quanta through the curvature barrier.

\begin{figure*}[tbhp]
	\centerline{
		\includegraphics[width=80mm,height=70mm]{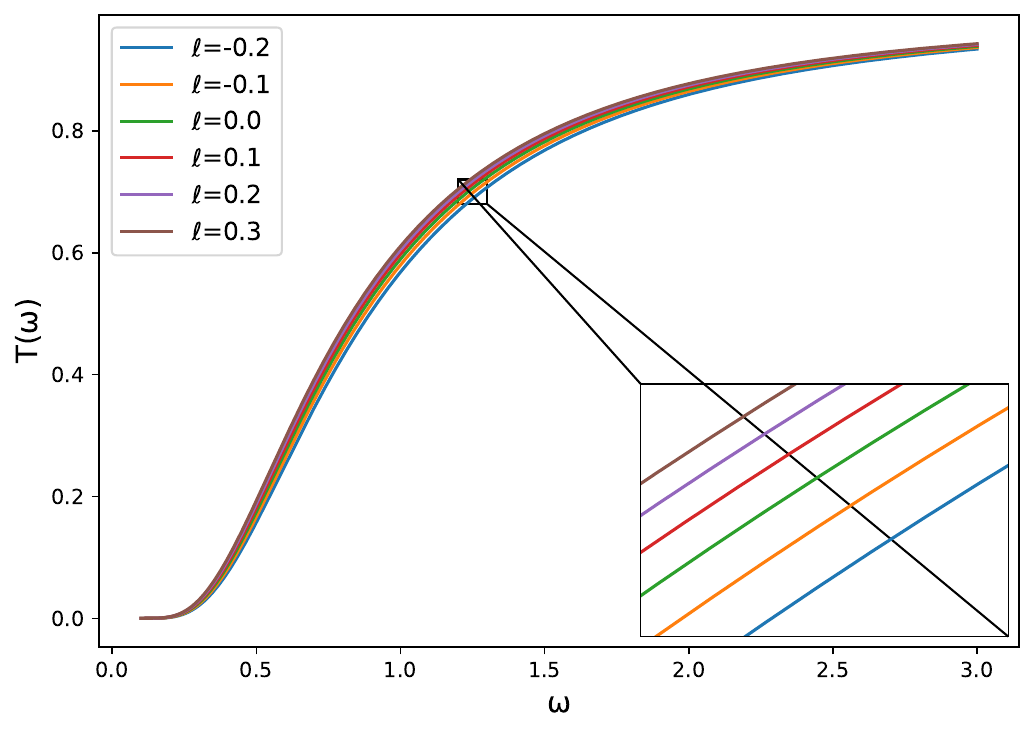}\qquad
        \includegraphics[width=80mm,height=70mm]{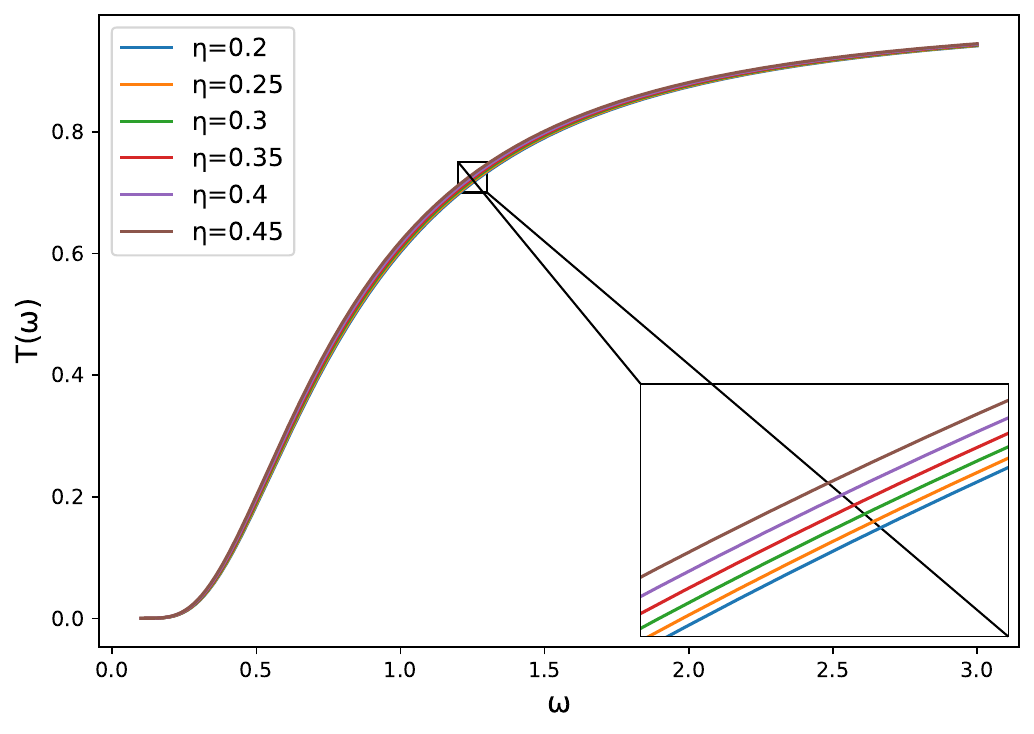}}(a) $\eta=0.25$ \hspace{5cm} (b) $\ell=0.2$
	\caption{Behavior of transmission coefficient as a function of frequency $\omega$ for fixed $\bar \ell=1$ and $Q=0.8$. Panel (a) corresponds to $\eta=0.25$ with varying $\ell$, while panel (b) corresponds to $\ell=0.2$ with varying $\eta$. }
\label{fig:GraybodyfacorTransm}
\end{figure*}

\begin{figure*}[tbhp]
	\centerline{
		\includegraphics[width=80mm,height=70mm]{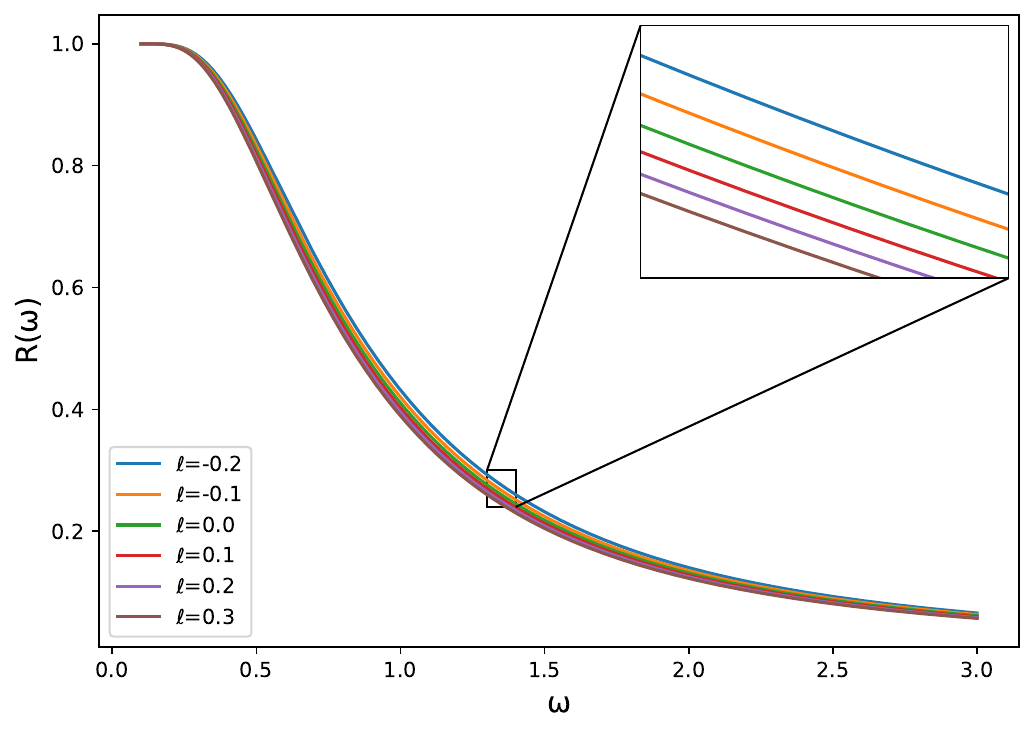}\qquad
        \includegraphics[width=80mm,height=70mm]{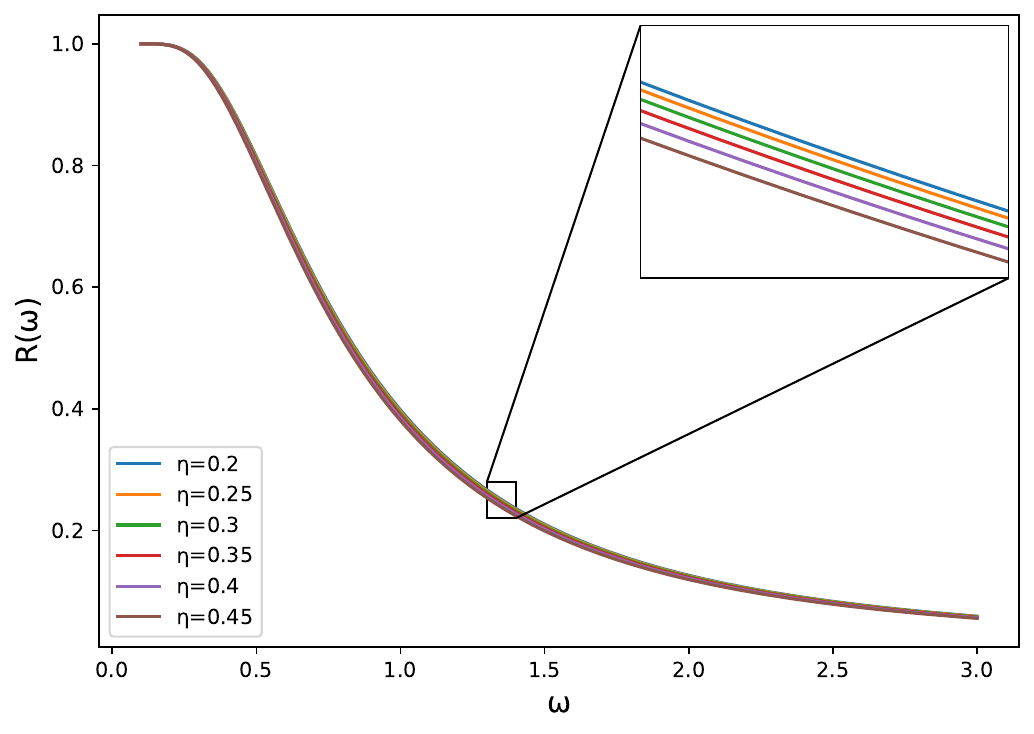}}(a) $\eta=0.25$ \hspace{5cm} (b) $\ell=0.2$
	\caption{Behavior of reflection coefficient as a function of frequency $\omega$ for fixed $\bar \ell=1$ and $Q=0.8$. Panel (a) corresponds to $\eta=0.25$ with varying $\ell$, while panel (b) corresponds to $\ell=0.2$ with varying $\eta$. }
\label{fig:GraybodyfacorRefl}
\end{figure*}
Figs.~\ref{fig:GraybodyfacorTransm}-\ref{fig:GraybodyfacorRefl} illustrates the behavior of the greybody factors, namely the transmission coefficient $T(\omega)$ and reflection coefficient $R(\omega)$, as functions of the frequency $\omega$ for different values of the LV parameter $\ell$ and the GM parameter $\eta$. In Fig.~\ref{fig:GraybodyfacorTransm}, the transmission coefficient $T(\omega)$ increases monotonically with frequency and approaches unity at high frequencies, indicating a transparent potential barrier. For fixed $\eta$, increasing $\ell$ suppresses transmission at low $\omega$, whereas larger $\eta$ enhances it. The inset highlights that these effects are most pronounced in the low-frequency regime.
In Fig.~\ref{fig:GraybodyfacorRefl}, the reflection coefficient $R(\omega)$ decreases with frequency and tends to zero at high $\omega$, consistent with $R+T=1$. For fixed $\eta$, increasing $\ell$ enhances reflection at low frequencies, while increasing $\eta$ reduces it. The zoomed region shows that parameter dependence is significant mainly at low and intermediate frequencies. In summary, the parameters $\ell$ and $\eta$ significantly influence the greybody factors by modifying the effective potential barrier experienced by the propagating field. Physically, $\ell$, associated with Lorentz-violating effects, tends to enhance the barrier strength, thereby suppressing transmission and increasing reflection at low frequencies. On the other hand, the GM parameter $\eta$ effectively reduces the barrier, increasing the transmission probability and correspondingly decreasing the reflection. As the frequency $\omega$ increases, the wave energy becomes sufficiently large to overcome the potential barrier, rendering the effects of both parameters negligible in the high-frequency regime.

\subsection{Energy Emission Rate}

In the high-frequency (geometric-optics) limit, the absorption cross section oscillates around a limiting constant value, denoted by $\sigma_{\rm lim}$ \cite{Sanchez1978,Decanini2011}. Since the capture of high-energy quanta is governed by null geodesics, the corresponding limiting absorption cross-section can be estimated from the black hole shadow as \cite{WeiLiu2013,LiEtAl2022}
\begin{equation}
\sigma_{\rm lim}\approx \pi R_{\rm sh}^{2},
\label{eer1}
\end{equation}
where $R_{\rm sh}$ is the shadow radius given by Eq.~(\ref{dd12}).

Within this approximation, the spectral energy emission rate is written as \cite{Page1976,WeiLiu2013,LiEtAl2022,Kala:2022uog}
\begin{equation}
\frac{d^{2}E}{d\omega\,dt}
=
\frac{2\pi^{2}\sigma_{\rm lim}}{e^{\omega/T_H}-1}\,\omega^{3},
\label{eer2}
\end{equation}
where $\omega$ is the emission frequency and $T_H$ is the Hawking temperature given by Eq.~(\ref{cc6}).

Substituting Eq.~(\ref{eer1}) into Eq.~(\ref{eer2}), we obtain
\begin{equation}
\frac{d^{2}E}{d\omega\,dt}
=
\frac{2\pi^{3}R_{\rm sh}^{2}\,\omega^{3}}{e^{\omega/T_H}-1}.
\label{eer3}
\end{equation}
Using Eqs.~(\ref{cc6}) and (\ref{dd12}), the above expression can be written explicitly as
\begin{equation}
\frac{d^{2}E}{d\omega\,dt}
=
\frac{2\pi^{3}\beta^{2}r_{s}^{2}\,\omega^{3}}
{
f(r_s)\left[
\exp\!\left(
\frac{4\pi r_h\sqrt{1+\ell}\,\omega}
{1-\varsigma Q^{2}/r_h^{2}}
\right)-1
\right]
},
\label{eer4}
\end{equation}
where $r_s$ is the photon-sphere radius given by Eq.~(\ref{dd8}).

Equation~(\ref{eer4}) shows that the energy emission rate is controlled by both the Hawking temperature and the shadow radius. Therefore, the Lorentz-violating parameter $\ell$, the global monopole parameter $\eta$, and the electric charge $Q$ modify the spectral profile of the emitted radiation. In particular, changes in these parameters may shift the emission peak and alter its amplitude, providing an additional observable signature of the geometry under consideration. Related analyses of shadow-controlled emission spectra in modified black-hole backgrounds can also be found in Refs.~\cite{WeiLiu2013,LiEtAl2022}. 

\begin{figure*}[tbhp]
    \centering
    \includegraphics[width=160mm,height=70mm]{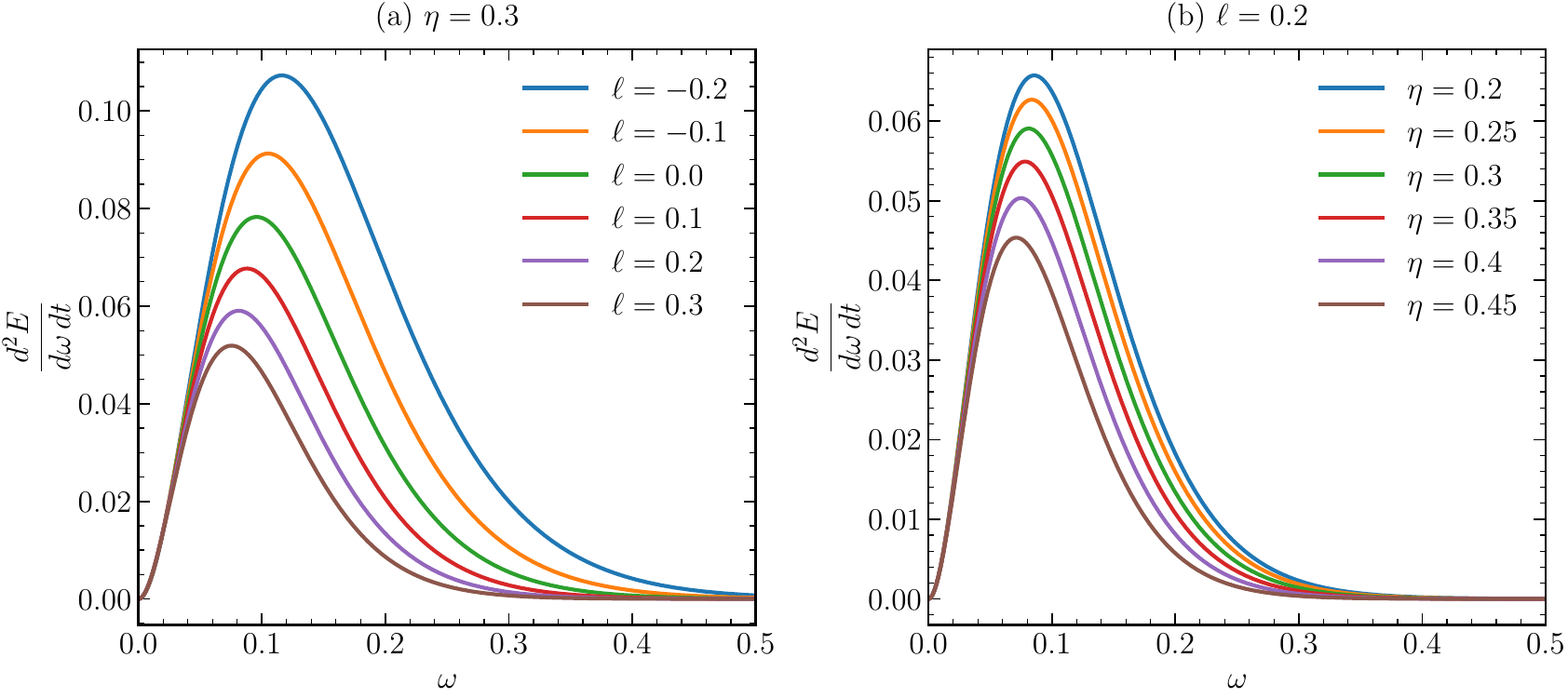}
    \caption{Energy emission rate as a function of the frequency $\omega$ for fixed $M=1$ and $Q=0.8$. Panel (a) corresponds to $\eta=0.3$ with varying $\ell$, while panel (b) corresponds to $\ell=0.2$ with varying $\eta$. The peak position and peak height encode the combined influence of the Hawking temperature and the shadow radius.}
    \label{fig:EnergyEmission}
\end{figure*}

Figure~\ref{fig:EnergyEmission} shows the behavior of the energy emission rate as a function of the frequency $\omega$ for different values of the parameters $\ell$ and $\eta$. In Fig.~\ref{fig:EnergyEmission}(a), the global monopole parameter is fixed at $\eta=0.3$ while the Lorentz-violating parameter $\ell$ varies. In Fig.~\ref{fig:EnergyEmission}(b), the Lorentz-violating parameter is fixed at $\ell=0.2$ while the global monopole parameter $\eta$ varies. In both panels, the emission spectrum exhibits the characteristic Planck-like profile: it starts from zero, rises to a maximum, and then decays exponentially at high frequencies. Variations in $\ell$ and $\eta$ modify both the height and the position of the peak through their influence on $T_H$ and $R_{\rm sh}$. In general, a reduction in the Hawking temperature suppresses the emission peak and shifts it toward lower frequencies. Thus, the Lorentz-violating and global monopole parameters leave measurable imprints on the black hole radiation spectrum.

\subsection{Sparsity of Hawking Radiation}

Although Hawking radiation possesses a thermal spectrum, the emission process is not continuous in time. Instead, it occurs as a sequence of well-separated quanta, which means that the radiation flux is intrinsically sparse \cite{VisserEtAl2017,ChowdhuryBanerjee2020}. A convenient measure of this effect is obtained by comparing the typical thermal wavelength of the emitted quanta with the effective emission area of the black hole \cite{VisserEtAl2017}.

We define the thermal wavelength as
\begin{equation}
\lambda_{\rm th}=\frac{2\pi}{T_H},
\label{sp1}
\end{equation}
and the effective area as
\begin{equation}
A_{\rm eff}=\frac{27}{4}A_H,
\label{sp2}
\end{equation}
where the horizon area is
\begin{equation}
A_H=4\pi\beta^{2}r_h^{2}.
\label{sp3}
\end{equation}

The sparsity parameter is then defined by \cite{VisserEtAl2017,ChowdhuryBanerjee2020}
\begin{equation}
\eta_{\rm sp}=\frac{\lambda_{\rm th}^{2}}{A_{\rm eff}}.
\label{sp4}
\end{equation}
Using Eqs.~(\ref{cc6}) and (\ref{sp2})--(\ref{sp4}), we find
\begin{equation}
\eta_{\rm sp}
=
\frac{64\pi^{3}}{27}\,
\frac{1+\ell}{\beta^{2}}
\left(
1-\varsigma\,\frac{Q^{2}}{r_h^{2}}
\right)^{-2}.
\label{sp5}
\end{equation}
Since $\beta^{2}=1-\eta^{2}$ and $\varsigma=1/(\lambda \beta^{4})$, the above expression becomes
\begin{equation}
\eta_{\rm sp}
=
\frac{64\pi^{3}}{27}\,
\frac{1+\ell}{1-\eta^{2}}
\left[
1-
\frac{2+\ell}{2(1+\ell)(1-\eta^{2})^{2}}
\frac{Q^{2}}{r_h^{2}}
\right]^{-2}.
\label{sp6}
\end{equation}

In the Schwarzschild limit, namely $Q=0$, $\ell=0$, and $\eta=0$, one recovers the standard result \cite{VisserEtAl2017}
\begin{equation}
\eta_{\rm sp}^{\rm Sch}=\frac{64\pi^{3}}{27}.
\label{sp7}
\end{equation}
Some useful limiting cases are
\begin{align}
\eta_{\rm sp}\big|_{Q=0}
&=
\frac{64\pi^{3}}{27}\,
\frac{1+\ell}{1-\eta^{2}},
\label{sp8}
\\
\eta_{\rm sp}\big|_{\eta=0}
&=
\frac{64\pi^{3}}{27}\,
(1+\ell)
\left[
1-\frac{2+\ell}{2(1+\ell)}\frac{Q^{2}}{r_h^{2}}
\right]^{-2},
\label{sp9}
\\
\eta_{\rm sp}\big|_{\ell=0}
&=
\frac{64\pi^{3}}{27}\,
\frac{1}{1-\eta^{2}}
\left[
1-\frac{Q^{2}}{(1-\eta^{2})^{2}r_h^{2}}
\right]^{-2}.
\label{sp10}
\end{align}

Equation~(\ref{sp6}) shows that the sparsity of Hawking radiation is directly affected by the Lorentz-violating parameter, the global monopole, and the electric charge. Larger values of $\eta_{\rm sp}$ correspond to a more sparse flux, meaning that the emitted quanta are more widely separated in time. Therefore, the geometry considered here may lead to substantial deviations from the standard Schwarzschild behavior. Similar applications of the sparsity analysis to nonstandard black-hole spacetimes can be found in Refs.~\cite{ChowdhuryBanerjee2020}.

\begin{figure*}[tbhp]
    \centering
    \includegraphics[width=160mm,height=70mm]{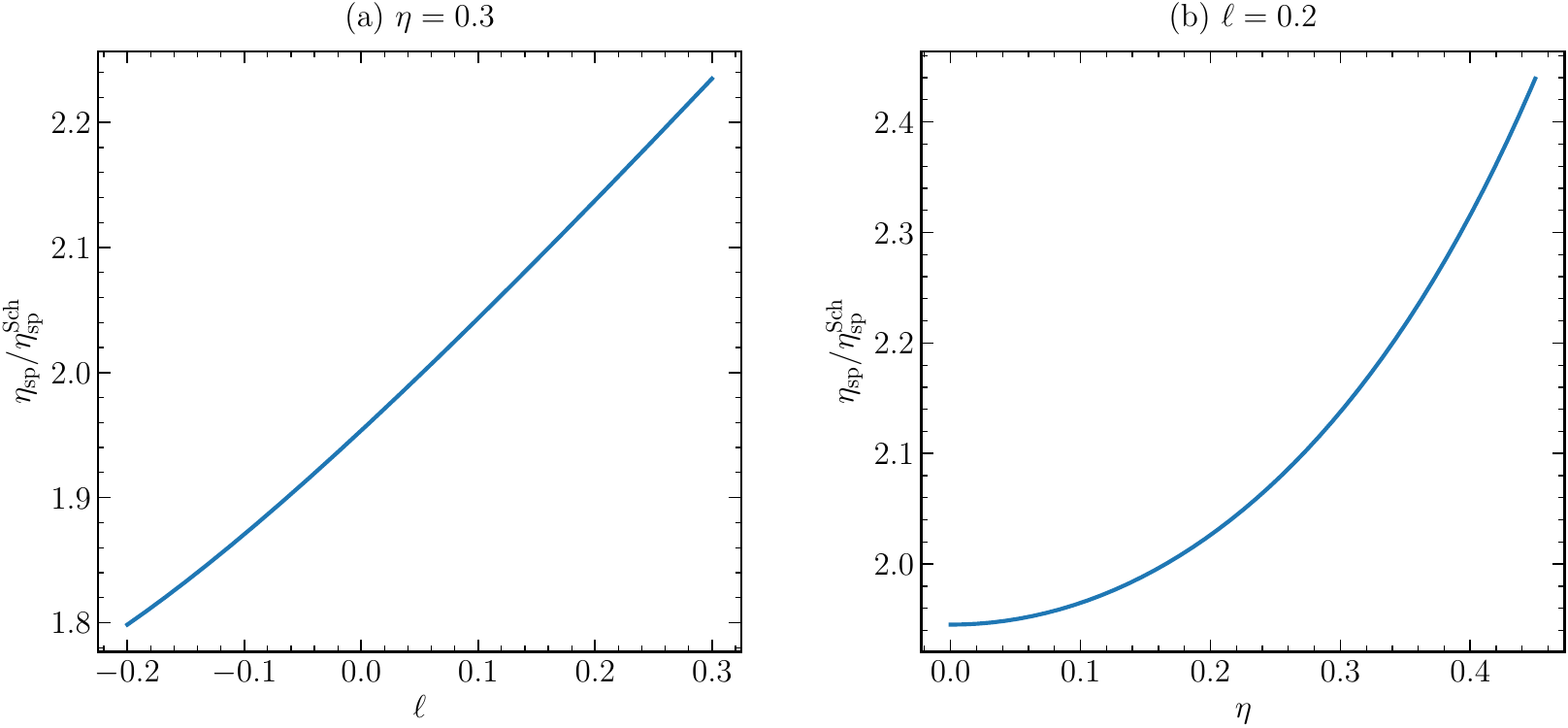}
    \caption{Normalized sparsity parameter $\eta_{\rm sp}/\eta_{\rm sp}^{\rm Sch}$ for fixed $M=1$ and $Q=0.8$. Panel (a) corresponds to $\eta=0.3$ with varying $\ell$, while panel (b) corresponds to $\ell=0.2$ with varying $\eta$. Larger values indicate a more sparse Hawking flux, i.e., a greater temporal separation between emitted quanta.}
    \label{fig:Sparsity}
\end{figure*}

Figure~\ref{fig:Sparsity} illustrates the behavior of the sparsity parameter for different values of $\ell$ and $\eta$. In Fig.~\ref{fig:Sparsity}(a), the global monopole parameter is fixed at $\eta=0.3$ while the Lorentz-violating parameter $\ell$ varies. In Fig.~\ref{fig:Sparsity}(b), the Lorentz-violating parameter is fixed at $\ell=0.2$ while the global monopole parameter $\eta$ varies. The results indicate that the Hawking flux becomes increasingly sparse as the geometric parameters move away from the Schwarzschild limit. This reflects the combined effects of the conical deficit, Lorentz violation, and electric charge on the thermal wavelength and effective emission area. Hence, the sparsity parameter provides another useful quantity for characterizing the observational properties of the black hole.

\section{Conclusions}\label{summary}

In this work, we investigated a charged black hole solution in bumblebee gravity in the presence of a global monopole, with emphasis on how Lorentz-violating effects, electric charge, and a conical deficit jointly modify the thermodynamic, optical, dynamical, and radiative properties of the system. Our analysis provides a unified picture of how these parameters reshape the strong-gravity phenomenology of a nonstandard charged black hole.

On the thermodynamic side, we derived the horizon structure, entropy, Hawking temperature, Gibbs free energy, electrostatic potential, and specific heat. The results show that both the Lorentz-violating parameter $\ell$ and the monopole parameter $\eta$ produce nontrivial deformations in the thermal sector. In particular, they shift the temperature peak, alter the Gibbs free-energy profile, and move the critical point at which the specific heat diverges. This indicates that the local thermal stability and phase structure of the black hole are sensitive to both Lorentz-violating geometry and topological defects.

In the optical sector, we obtained the photon-sphere radius, the critical impact parameter, and the shadow radius in the asymptotically conical spacetime. We showed that increasing $\ell$ or $\eta$ generally enlarges the photon sphere and the apparent shadow size. We also discussed how the EHT bounds on the shadow of Sgr~A* can be used to constrain the parameter space of the model. From this perspective, the shadow remains one of the most direct observational probes of the combined effects of charge, Lorentz violation, and global monopoles.

We further studied null and timelike geodesics in both the weak-field and strong-field regimes. For photon trajectories, the weak deflection angle acquires not only the usual mass and charge corrections, but also a topological contribution associated with the conical asymptotics. For massive particles, we derived the corrected perihelion shift and showed explicitly that both $\ell$ and $\eta$ modify the precession of bound orbits. In addition, our analysis of neutral circular motion revealed how the ISCO-related quantities, the specific angular momentum, the specific energy, and the azimuthal angular velocity are shifted by the geometric parameters.

In the perturbative sector, we derived the effective potential for a massless scalar field, obtained semi-analytic bounds on the greybody factors, and then used the corrected Hawking temperature and shadow radius to study the energy emission rate and the sparsity of Hawking radiation. In addition, QNMs exhibit a decrease in the real part of the frequency as the LV and GM parameters increase, indicating a suppression of the oscillation frequency. Meanwhile, the imaginary part shows a non-monotonic dependence on both parameters, revealing a complex influence on the damping rate and hence the stability of the black hole perturbations. These quantities provide a complementary radiative characterization of the black hole. In particular, the energy-emission spectrum is sensitive to the same geometric parameters that govern the shadow, while the sparsity parameter quantifies how far the Hawking flux deviates from a continuous thermal picture. 

Overall, our results show that the charged black hole in bumblebee gravity with a global monopole provides a rich and internally consistent arena in which thermodynamics, lensing, orbital dynamics, perturbation theory, and Hawking radiation are all modified in correlated ways. The Lorentz-violating parameter $\ell$, the monopole parameter $\eta$, and the charge $Q$ do not merely deform the metric quantitatively; together they produce qualitative changes in the location of characteristic radii, in the stability of thermal branches, in the strength of weak and strong gravitational lensing, and in the properties of emitted radiation.

These features make the present solution an interesting candidate for future extensions. Possible directions include rotating generalizations, quasinormal-mode analyses, more detailed observational comparisons with black hole imaging data, and a broader study of Hawking radiation beyond the semi-analytic bounds considered here. We hope that the present work will serve as a useful step toward a more complete understanding of Lorentz-violating and topologically nontrivial black holes in strong-field gravity.

\footnotesize

\section*{Acknowledgements}

F.A. acknowledges the Inter University Centre for Astronomy and Astrophysics (IUCAA), Pune, India, for granting a visiting associateship. S.K. sincerely acknowledges IMSc for providing exceptional research facilities and a conducive environment that facilitated his work as an Institute Postdoctoral Fellow. E. O. Silva acknowledges the support from Conselho Nacional de Desenvolvimento Cient\'{i}fico e Tecnol\'{o}gico (CNPq) (grants 306308/2022-3), Funda\c c\~ao de Amparo \`{a} Pesquisa e ao Desenvolvimento Cient\'{i}fico e Tecnol\'{o}gico do Maranh\~ao (FAPEMA) (grants UNIVERSAL-06395/22), and Coordena\c c\~ao de Aperfei\c coamento de Pessoal de N\'{i}vel Superior (CAPES) - Brazil (Code 001).

\end{document}